\documentclass[seceq]{ptptex}

\usepackage{graphicx}
\usepackage{wrapft}

\renewcommand{\a}{\alpha}
\renewcommand{\b}{\beta}
\renewcommand{\c}{\gamma}
\renewcommand{\d}{\delta}
\newcommand{\e}{\epsilon}

\newcommand{\m}{\mu}
\newcommand{\n}{\nu}
\renewcommand{\t}{\tau}
\newcommand{\z}{\omega}
\newcommand{\G}{\Gamma}

\newcommand{\ol}{\overline}

\title{
New Numerical Methods to Evaluate Homogeneous Solutions of 
the Teukolsky Equation
}

\author{
Ryuichi \textsc{Fujita} and Hideyuki \textsc{Tagoshi}%
}

\inst{
Department of Earth and Space Science, Graduate School of Science,\\
Osaka University, Toyonaka 560-0043, Japan\\
}

\abst{
We discuss a numerical method to compute the homogeneous solutions of the
Teukolsky equation which is the basic equation of the black hole perturbation
method. We use the formalism developed by 
Mano, Suzuki and Takasugi, in which the homogeneous solutions of the radial Teukolsky 
equation are expressed in terms of two kinds of series of special functions, and the
formulas for the asymptotic amplitudes are derived explicitly.
Although the application of this method was previously limited to the analytical 
evaluation of the homogeneous solutions, 
we find that it is also useful for numerical computation. 
We also find that so-called "renormalized angular momentum parameter", $\nu$, 
can be found only in the limited region of $\omega$ for each $l,m$
if we assume $\nu$ is real (here, $\omega$ is the angular frequency, and 
$l$ and $m$ are degree and order of the spin-weighted spheroidal harmonics
respectively). 
We also compute the flux of the gravitational waves induced by a compact star
in a circular orbit on the equatorial plane around a rotating black hole. 
We find that the relative error of the energy flux is about $10^{-14}$
which is much smaller than the one 
obtained by usual numerical integration methods. 
}

\begin{document}

\maketitle

\section{Introduction}
Inspirals of stellar-mass compact objects into a supermassive 
black hole at galactic nuclei are expected to be one of the most important 
sources of the gravitational waves for space-based detectors, such 
as the Laser Interferometer Space Antenna (LISA) \cite{LISA}. 
Current best estimate of the number of such event is given by Gair et al.\cite{Gair}. 
They estimated the number of event for the inspirals of 
$10M_\odot$ black holes into $10^6M_\odot$ suppermassive black holes 
to be 660 by 3 years observation. 
By observing gravitational waves from such systems, we may be able to obtain
information of the central black hole's spacetime geometry encoded in 
multipole moments, and test the validity of the no-hair theorem of 
black hole\cite{Ryan}. 
We may also obtain astrophysical information such like 
the mass distribution of compact objects in galactic nuclei.
The optimal detection strategy for such gravitational waves 
is matched filtering, which requires theoretical 
waveforms to be correlated with the data. 
Although we may not need very accurate waveforms in detection, 
we will need very accurate theoretical waveforms to 
extract astrophysical information concerning the source.

To predict the waveforms of extreme mass ratio inspirals, we adopt the 
black hole perturbation approach. In this approach, 
a compact star is treated as a point particle, and the mass of the compact star, $\mu$,
is assumed to be very small compared to the mass of 
the black hole, $M$, i.e. $\mu/M\ll 1$. 
In this context, the Teukolsky equation \cite{Teukolsky}
describes the evolution of a perturbation of the Kerr black hole 
spacetime. The standard approach to solve the Teukolsky equation is 
based on the Green function method. The Green function is constructed 
using two kinds of homogeneous solutions. The solution of the 
Teukolsky equation is obtained by integrating the Green function
multiplied by the source term. 
In the case of a Kerr black hole, 
the homogeneous solution is calculated usually 
by the Sasaki-Nakamura equation, \cite{SN}
which is derived with the Sasaki-Nakamura transformation from the Teukolsky equation. 
The Sasaki-Nakamura transformation is a generalization of the Chandrasekhar
transformation \cite{Chandra} by which we can obtain the Regge-Wheeler equation from 
the Bardeen-Press-Teukolsky equation \cite{Teukolsky, BP}in the Schwarzschild case. 
The Sasaki-Nakamura equation is also a powerful formula when we compute
the gravitational wave flux induced by particles in unbound orbits. 

In the past, there were several works which calculated the flux and 
waveforms of gravitational waves induced by a compact star around a black hole
computed by the Teukolsky and the Sasaki-Nakamura equations, 
and its effects to the orbital evolution of the star
under the influence of radiation reaction. 
Here, we list some of recent works treating bound orbits 
of the star. 
Many other works can also be 
found in a review article by Nakamura et al.\cite{NOK}

Shibata\cite{Shibata1} calculated the gravitational waves 
in the case of circular, equatorial orbits around a Kerr black hole.
Finn and Thorne\cite{FT} consider the same case, and discussed 
the signal-to-noise ratio of such gravitational waves 
when detected by LISA and the detectability of such sources by LISA. 
Tanaka et al.\cite{TSSTT} and Cutler et al.\cite{CKP}
calculated the gravitational waves in the case of eccentric orbits 
around a Schwarzschild black hole, and discussed orbital 
evolution of the star. 
Apostolatos et al.\cite{AKOP} consider the slightly eccentric orbits
around a Schwarzschild black hole, and discussed the stability of the circular
orbit under the influence of radiation reaction. 
Kennefick\cite{Kennefick} consider the slightly orbits on the equatorial plane 
around a Kerr black hole, and discussed the stability of the circular orbit 
under the influence of radiation reaction. 
Shibata\cite{Shibata2} calculated the gravitational waves in the case of 
eccentric, equatorial orbits around a Kerr black hole, and discuss the 
importance of the black hole spin and other relativistic effects. 
Glampedakis and Kennefick\cite{GK} calculated the gravitational waves in the case of 
eccentric, equatorial orbits around a Kerr black hole, and discuss the 
orbital evolution of the star under the influence of radiation reaction. 
Shibata\cite{Shibata3} calculated the gravitational waves in the case of circular,
nonequatorial orbits around a Kerr black hole, and discussed the 
radiation reaction effects. 
Hughes\cite{Hughes1,Hughes2} calculated the gravitational waves 
in the case of circular, nonequatorial orbits around a Kerr black hole, and 
discuss the radiation reaction effects. 

The source term of the Teukolsky equation 
is obtained by specifying the orbit of the particle. 
In the case of a Kerr black hole, orbits around it are specified by 
the energy, the $z$-component of the angular momentum, and the Carter constant.
In the case of bound orbits, 
when the orbit is limited to the equatorial plane but is not circular,
the orbits show exhibit the "zoom-whirl" property as the eccentricity becomes larger
\cite{GK}. When the orbit is no longer limited to the equatorial plane, 
which may be important for sources of LISA, 
the orbits become more complicated. 
In such cases, we have to trace the orbit for much longer time than the dynamical 
time of the system in order to integrate 
the source term multiplied by the Green function accurately. 
Further, many side-bands in the spectrum of gravitational waves
must be calculated to establish the good accuracy. 

Let us assume for simplicity that LISA will observe gravitational waves 
at the frequency $f\sim 10^{-2}$Hz for one year. 
The total cycle of waves is typically $N_{\rm cycle}\sim 10^{5}$. 
Thus, the relative error for the luminosity $\Delta\dot{E}/\dot{E}$ required to establish 
the accuracy for the cycle, $\Delta N_{\rm cycle}\leq 1$ is 
$\Delta \dot{E}/\dot{E}\leq 10^{-5}$ \cite{CFPS} in the simplest, circular 
orbit cases. For more complicated orbits, the requirement to the accuracy 
would be stronger than this. 
Although the accuracy, $10^{-5}$, is already established in many of previous works, 
it would be very helpful for the future data analysis of LISA
if we had more efficient and accurate methods to 
compute the homogeneous solutions of the Teukolsky equation
to calculate the gravitational wave flux. 

Among approaches to obtain homogeneous solutions of the
Teukolsky equation, Leaver \cite{Leaver} formulated a method to express them
in terms of a series of Coulomb wave functions. 
Although this series is convergent 
at the spatial infinity, it does not converge around the horizon. 
For this reason, it was suggested that a power series expansion around
the horizon be used 
to obtain all of the asymptotic amplitudes of the homogeneous solution 
which are needed in the Green function. 
Using this method, Tagoshi and Nakamura \cite{TN}
carried out a high precision computation of a homogeneous solution of 
the Teukolsky equation and obtained high precision numerical data of the gravitational
wave flux induced by a particle in a circular orbit around a Schwarzschild black hole. 
In this calculation, the convergence of the series of Coulomb wave functions 
is very fast, and we can evaluate the series of Coulomb wave functions
very accurately. Contrastingly, because the convergence of the power series expansion 
is very slow, it is not easy to evaluate it very accurately. 
In fact, the accuracy of the numerical data of the homogeneous solution is limited by 
the accuracy of the power series expansion.

Later, Mano, Suzuki and Takasugi (MST) \cite{MST} formulated a method to express a 
homogeneous solution in a series of hypergeometric functions around the 
horizon. This series has a very convenient property that allows it to
be matched with the series of Coulomb wave 
functions around the infinity. Specifically, this property is that the three term recurrence
relation among the expansion coefficients for the hypergeometric case 
is the same as that in Coulomb case. Owing to this property, it is possible 
to match the two series expansions analytically, and we can 
express the asymptotic amplitude of the homogeneous solution in terms
of only 
the expansion coefficients, not 
in terms of the hypergeometric or Coulomb functions themselves. 
This property is that which distinguishes the MST method from the Leaver method. 
Because we do not need to use a power series expansion 
to determine the asymptotic amplitudes by numerical matching, 
it is expected that the matching can be done very efficiently in the MST method.

So far, the MST method has been used only in analytic calculations. 
There is a close relation between the Coulomb or hypergeometric
series expansion and the low frequency expansion of the Teukolsky equation. 
Here, low frequency usually implies small post-Minkowskian or post-Newtonian expansion
parameters. We can calculate the higher order terms of the 
post-Newtonian expansion of the gravitational wave flux from a black hole \cite{chapter,ST}
systematically using this method. 
Tagoshi, Mano and Takasugi \cite{TMT} also 
computed the energy absorbed by a rotating black hole analytically 
assuming that the orbit of the particle is very large. 

In this paper, we use the MST formalism for numerical computation of the 
homogeneous Teukolsky equation. 
We discuss the numerical method to calculate the homogeneous solutions
in detail. One of the most important problem involving the MST formalism 
is to determine the so-called "renormalized angular momentum" $\nu$ by solving
the equation $g(\nu)=0$, which is expressed in terms of continued fractions. 
We investigate the numerical properties of this function $g(\nu)$. 
We find that, for each $s, \ell, m$ and $q$ ($s$ is the spin index of the Teukolsky 
equation, $\ell,m$ are the indices of the
spin weighted spheroidal harmonics, and $q$ is the Kerr parameter divided by 
the mass of black hole, $q=a/M$.), there is a maximum value of
$M\omega$ for which
we can find $\nu$, assuming $\nu$ to be real. 
Because the applicability of the MST formalism depends on the existence of $\nu$, 
this is a serious problem. 
Although our preliminary investigation suggests that we can find $\nu$ if we assume 
it to be complex, more effort is needed to establish the numerical accuracy. 
Therefore, we restrict the region of $M\omega$ for which we can find real $\nu$. 

Once we have the renormalized angular momentum $\nu$, it is straightforward to evaluate 
the expansion coefficients and the series of hypergeometric functions or Coulomb wave 
functions. As expected, we find that the convergence of the series of hypergeometric functions
and Coulomb wave functions is very fast, and we can evaluate the homogeneous solutions 
very accurately. 

As a test calculation, we compute the gravitational wave flux
induced by a particle in a circular orbit on the equatorial plane
around a Kerr black hole. This is because that the numerical data 
for the energy flux can easily be compared with the one obtained by other methods. 
The accuracy of the numerical data is compared 
with that of previous works. 

This paper is organized as follows. 
In $\S$ \ref{sec:MST}, we review the Teukolsky formalism and 
the MST method. In $\S$ \ref{sec:NM}, we 
discuss the numerical method to calculate the homogeneous solutions. 
We present numerical results of the gravitational wave flux induced 
by a particle in circular, equatorial orbits around a Kerr black hole 
in $\S$ \ref{sec:luminosity}. 
$\S$ \ref{sec:summary} is devoted to the summary and discussion. 
Throughout this paper, we use units in which $G=c=1$.

\section{Analytic solutions of the homogeneous Teukolsky equation}
\label{sec:MST}
\subsection{The homogeneous solutions in series of hypergeometric functions and 
Coulomb wave functions}

The radial Teukolsky equation is given by (see Appendix \ref{sec:Teukolsky})
\begin{eqnarray}
\label{eq:MSTeukolsky}
\Delta^2{d\over dr}\left({1\over \Delta}{dR_{\ell m\omega}\over dr}
\right)
-V(r) R_{\ell m\omega}=T_{\ell m\omega},
\end{eqnarray}
where the potential term $V(r)$ is given by
\begin{eqnarray}
V(r) = -{K^2 + 4i(r-M)K\over\Delta} + 8i\omega r + \lambda.
\end{eqnarray}
Here $\Delta=r^2-2Mr+a^2=(r-r_+)(r-r_-)$ with $r_\pm=M\pm\sqrt{M^2-a^2}$, $K=(r^2+a^2)\omega-ma$ and $\lambda$ is the 
eigenvalue of the angular Teukolsky equation.

In the MST method, the homogeneous solutions of the Teukolsky equation 
are expressed in terms of two kinds of series of special 
functions \cite{MST,MSTR}. One consists of series of hypergeometric 
functions, and the other consists of series of Coulomb wave functions.
The former is convergent at the horizon and the latter at infinity. 
We match the two kinds of solutions in the overlapping region of convergence. 
We thereby obtain analytical expressions of the asymptotic amplitudes of the solution.
(See Ref. \citen{ST} for a more recent review.)

First, we present a solution consisting of a series of hypergeometric functions. 
For the incoming solution $R_{lm\omega}^{{\rm in}}$, we define
$p_{{\rm in}}$ by 
\begin{equation}
R_{lm\omega}^{{\rm in}}=e^{i\e\kappa x}(-x)^{-s-i(\e+\t)/2}
(1-x)^{i(\e-\t)/2}p_{{\rm in}}(x).
\label{eq:Rin}
\end{equation}
The function $p_{\rm in}$ is expressed in a series of hypergeometric
functions as 
\begin{eqnarray}
\label{eq:series of Rin}
p_{{\rm in}}(x)=\displaystyle\sum_{n=-\infty}^{\infty}a_{n}
F(n+\nu+1-i\tau,-n-\n-i\tau;1-s-i\e-i\tau;x),
\end{eqnarray}
where 
$x=\z (r_+ -r)/\e \kappa$, $\e=2M\z, \kappa={\sqrt{1-q^2}}, q=a/M, 
\t={{(\e-mq)}/{\kappa}}$ and $F(\a,\b;\c;x)$ is the hypergeometric function.

This expression can be rewritten, using the analytic properties of hypergeometric
functions, into the form of a series expansion with better convergence properties for large $|x|$ as 
\begin{equation}
R_{lm\omega}^{\rm in}=R^\nu_0+R^{-\nu-1}_0,
\end{equation}
where
\begin{eqnarray}
R_0^{\nu}&=&e^{i\epsilon \kappa x}(-x)^{-s-{i\over{2}}(\epsilon+\tau)}
(1-x)^{{i\over{2}}(\epsilon+\tau)+\nu}
\nonumber\\
&&
\times\sum_{n=-\infty}^{\infty}
f_n^{\nu}\,{\Gamma(1-s-i\epsilon-i\tau)\Gamma(2n+2\nu+1)
\over{\Gamma(n+\nu+1-i\tau)\Gamma(n+\nu+1-s-i\epsilon)}}
\nonumber\\
&\times&
(1-x)^{n}F(-n-\nu-i\tau,-n-\nu-s-i\epsilon;-2n-2\nu;{1\over{1-x}})\,.
\end{eqnarray}

Next, we present a solution in the form of series of Coulomb wave functions. 
Let us denote a Teukolsky function by $R_{{\rm C}}$. 
We define the function $f_{\n}(z)$ through the relation
\begin{equation}
R_{{\rm C}}={z}^{-1-s}\left(1-{\e \kappa \over{{z}}}\right)^{-s-i(\e+\t)/2}
f_{\n}(z). 
\end{equation}
The function $f_{\nu}(z)$ is expressed in a series of Coulomb wave functions as 
\begin{eqnarray}
\label{eq:series of Rc}
f_{\n}(z)= 
\displaystyle\sum_{n=-\infty}^{\infty}
(-i)^n\frac{(\n+1+s-i\e)_n}{(\n+1-s+i\e)_n}a_n F_{n+\n}(-is-\e,z),
\end{eqnarray}
where $z=\z (r- r_- )$, $(a)_{n}=\Gamma(a+n)/\Gamma(a)$
and $F_{N}(\eta,z)$ is a Coulomb wave function defined by 
\begin{eqnarray}
F_{N}(\eta,z)=e^{-iz}2^{N}z^{N+1}\frac{\G(N+1-i\eta)}{\G(2N+2)}\Phi(N+1-i\eta,
2N+2;2iz).
\label{eq:defcoulomb}
\end{eqnarray}
Here $\Phi(\a,\b;z)$ is the confluent hypergeometric function, 
which is regular at $z=0$ (see $\S$ 13 of Ref.~\citen{handbook}). 

In this method, the expansion coefficients 
of the series of hypergeometric functions and the series of Coulomb
wave functions $\{a_n\}$ exhibit the same recurrence relation. 
We find that the expansion coefficients $a_n$
satisfy the three-term recurrence relation
\begin{eqnarray}
\label{eq:3term}
\alpha_n^\nu a_{n+1}+\beta_n^{\nu} a_{n}+\gamma_n^\nu a_{n-1}=0,
\end{eqnarray}
where
\begin{eqnarray}
\a_n^\n&=&{i\e \kappa (n+\n+1+s+i\e)(n+\n+1+s-i\e)(n+\n+1+i\t)
\over{(n+\n+1)(2n+2\n+3)}},
\nonumber\\
\b_n^\n&=&-\lambda-s(s+1)+(n+\n)(n+\n+1)+\e^2+\e(\e-mq)
+{\e (\e-mq)(s^2+\e^2) \over{(n+\n)(n+\n+1)}},\nonumber\\
\c_n^\n&=&-{i\e \kappa (n+\n-s+i\e)(n+\n-s-i\e)(n+\n-i\t)
\over{(n+\n)(2n+2\n-1)}}.\nonumber
\end{eqnarray}
We note that the parameter $\nu$ introduced in the above formulas
does not exist in the Teukolsky equation. 
This parameter is introduced so that both series 
converge and actually represent a solution of the Teukolsky equation. 
We next introduce the following quantities:
\begin{equation}
R_n\equiv {a_{n}\over a_{n-1}}\,,\quad
L_n\equiv {a_{n}\over a_{n+1}}\,.
\end{equation}
We can express $R_{n}$ and $L_{n}$ in terms of continued fractions as 
\begin{eqnarray}
R_n&=&-{\gamma_n^\nu\over {\beta_n^\nu+\alpha_n^\nu R_{n+1}}}
\nonumber\\
&=&-{\gamma_{n}^\nu\over \beta_{n}^\nu-}\,
{\alpha_{n}^\n\gamma_{n+1}^\nu\over \beta_{n+1}^\nu-}\,
{\alpha_{n+1}^\n\gamma_{n+2}^\nu\over \beta_{n+2}^\nu-}\cdots,
\label{eq:Rncont}\\
L_n&=&-{\alpha_n^\nu\over {\beta_n^\nu+\gamma_n^\nu L_{n-1}}}
\nonumber\\
&=&-{\alpha_{n}^\nu\over \beta_{n}^\nu-}\,
{\alpha_{n-1}^\n\gamma_{n}^\nu\over \beta_{n-1}^\nu-}\,
{\alpha_{n-2}^\n\gamma_{n-1}^\nu\over \beta_{n-2}^\nu-}\cdots.
\label{eq:Lncont}
\end{eqnarray}
The expressions for $R_{n}$ and $L_{n}$ are valid if these continued 
fractions converge. 
It has been proved \cite{Gautschi}
that the continued fraction of the right-hand side of Eq. (\ref{eq:Rncont}) 
converges if and only if the 
recurrence relations Eq. (\ref{eq:3term}) 
possess a minimal solution as $n\rightarrow\infty$. 
A similar theorem can be proven regarding the converge as $n\rightarrow -\infty$
of the right-hand side of Eq. (\ref{eq:Lncont}). 
Because the recurrence relation Eq. (\ref{eq:3term}) possess minimal solutions
as $n\rightarrow \pm\infty$, the continued fractions on the right-hand sides
of Eqs. (\ref{eq:Rncont}) and (\ref{eq:Lncont}) converge. 
Although minimal solutions in the limits $n\rightarrow \infty$ and $n\rightarrow -\infty$ 
do not coincide in general, we can match them by appropriately choosing $\nu$. 
Suppose $\{f_n^\nu\}$ is a solution that is minimal
for both $n\rightarrow\pm\infty$. 
It is proved\cite{Gautschi} that 
the following relations are satisfied:
\begin{eqnarray}
\tilde{R}_n\equiv \frac{f_n^\nu}{f_{n-1}^\nu}
&=&-{\gamma_{n}^\nu\over \beta_{n}^\nu-}\,
{\alpha_{n}^\n\gamma_{n+1}^\nu\over \beta_{n+1}^\nu-}\,
{\alpha_{n+1}^\n\gamma_{n+2}^\nu\over \beta_{n+2}^\nu-}\cdots,
\label{eq:tildeRn}\\
\tilde{L}_n^\nu\equiv \frac{f_n^\nu}{f_{n+1}^\nu}
&=&-{\alpha_{n}^\nu\over \beta_{n}^\nu-}\,
{\alpha_{n-1}^\n\gamma_{n}^\nu\over \beta_{n-1}^\nu-}\,
{\alpha_{n-2}^\n\gamma_{n-1}^\nu\over \beta_{n-2}^\nu-}\cdots.
\label{eq:tildeLn}
\end{eqnarray}
This implies the relation
\begin{eqnarray}
\label{eq:consistency}
\tilde{R}_{n}\tilde{L}_{n-1}=1.
\end{eqnarray}
If we choose $\nu$ such that it satisfies the implicit equation for $\nu$, 
Eq. (\ref{eq:consistency}), for a certain $n$, we can obtain a minimal solution
$\{f_n^\nu\}$ that is valid over the entire range $-\infty<n<\infty$. 
For the minimal solution, ${f_n^\nu}$, we have
\begin{equation}
\lim_{n\rightarrow\infty}n\frac{f_n^\nu}{f_{n-1}^\nu}=\frac{i\epsilon\kappa}{2},
\quad
\lim_{n\rightarrow-\infty}n\frac{f_n^\nu}{f_{n+1}^\nu}=-\frac{i\epsilon\kappa}{2}.
\label{eq:asymptotics of minimal solution}
\end{equation}

The minimal solution is important for the convergence of the series 
Eq. (\ref{eq:series of Rin}) and Eq. (\ref{eq:series of Rc}). 
It can be proved that if we use the minimal solution $\{f_n^\nu\}$ 
for the expansion coefficients $\{a_n\}$, the series of hypergeometric
functions Eq. (\ref{eq:series of Rin}) converges for $x$ in the range 
$-\infty< x\leq 0$. (In fact, this is true for all complex values of $x$, except at 
$|x|=\infty$.)
It has also been proved that if the expansion coefficients are given by 
the minimal solution, the series Eq. (\ref{eq:series of Rc}) converges
for $z>\epsilon\kappa$ or, equivalently, $r>r_+$. 

Instead of Eq. (\ref{eq:consistency}), 
we can use an equivalent but practically more convenient 
form of an equation that determines the value of $\n$. Dividing 
Eq. (\ref{eq:3term}) by $a_{n}$, we find 
\begin{eqnarray}
\label{eq:determine_nu}
\beta_n^{\nu}+\alpha_n^\nu \tilde{R}_{n+1}+\gamma_n^\nu \tilde{L}_{n-1}=0,
\end{eqnarray}
where $R_{n+1}$ and $L_{n-1}$ are given by the continued
fractions Eq. (\ref{eq:tildeRn}) and Eq. (\ref{eq:tildeLn}) respectively. 

\subsection{Matching and the asymptotic amplitudes}

Now, we match the two kinds of solutions, $R_{0}^{\n}$ and $R_{{\rm C}}^{\n}$. 
If we expand solutions in powers of $\tilde x=1-x 
\equiv z/(\e\kappa)$, we see that both solutions behave like $\tilde x^{\n}$ 
multiplied by a single valued function of $\tilde x$ for large $\mid \tilde 
x\mid$, as long as $\omega >0$. 
Thus, if we assume $\omega >0$,
the analytic properties of $R_{0}^{\n}$ and $R_{{\rm C}}^{\n}$ 
are the same. That implies that these two solutions are identical up to a 
constant multiple. We set
\begin{eqnarray}
R_{0}^{\n}=K_{\n}R_{{\rm C}}^{\n}.
\end{eqnarray}
Then, by comparing each power of $\tilde x$ in the region where both solutions 
converge, i.e. $1\ll\mid \tilde x\mid<\infty$, we find
\begin{eqnarray}
K_{\n}
&=&\frac{e^{i\e\kappa}(2\e\kappa)^{s-\n-N}2^{-s}i^{N}\Gamma(1-s-i\e-i\tau)\Gamma(N+2\n+2)}
{\Gamma(N+\n+1-s+i\e)\Gamma(N+\n+1+i\tau)\Gamma(N+\n+1+s+i\e)}\nonumber \\
&&\times\left(\sum_{n=N}^{\infty}(-1)^{n}\frac{\Gamma(n+N+2\n+1)}{(n-N)!}
\frac{\Gamma(n+\n+1+s+i\e)\Gamma(n+\n+1+i\tau)}
{\Gamma(n+\n+1-s-i\e)\Gamma(n+\n+1-i\tau)}f_{n}^{\n}\right)\nonumber \\
&&\times\left(\sum_{n=-\infty}^{N}\frac{(-1)^{n}}{(N-n)!(N+2\n+2)_{n}}
\frac{(\n+1+s-i\e)_{n}}{(\n+1-s+i\e)_{n}}f_{n}^{\n}\right)^{-1},
\end{eqnarray}
where $N$ can be any integer, and the factor $K_{\nu}$ should be independent 
of the choice of $N$.

Now, we have an alternative expression for the ingoing-wave function 
$R_{lm\omega}^{{\rm in}}$,
\begin{eqnarray}
\label{eq:secondRin}
R_{lm\omega}^{{\rm in}}=K_{\n}R_{{\rm C}}^{\n}+K_{-\n-1}R_{{\rm C}}^{-\n-1},
\end{eqnarray}
which converges for $r>r_{+}$, including $r=\infty$. Combining 
Eq. (\ref{eq:secondRin}) with Eq. (\ref{eq:series of Rin}), 
which converges everywhere 
except at $r=\infty$, we have a complete set of 
analytic solutions for the ingoing-wave 
function.

Now, we can obtain analytical expressions for the asymptotic amplitudes 
$B^{{\rm trans}}_{lm\omega}$, 
$B^{{\rm inc}}_{lm\omega}$ and $B^{{\rm ref}}_{lm\omega}$, 
of $R_{lm\omega}^{{\rm in}}$ defined in Eq. (\ref{eq:Rin-asymp}).
Comparing $R_{lm\omega}^{{\rm in}}$ with Eq. (\ref{eq:series of Rin}) 
in the limit of $r\rightarrow r_{+}$ and Eq. (\ref{eq:secondRin}) in
the limit of $r\rightarrow\infty$, 
we find 
\begin{eqnarray}
B^{{\rm trans}}_{lm\omega}
&=&\left(\frac{\e\kappa}{\omega}\right)^{2s}e^{i\frac{\e+\t}
{2}\ln\kappa}
\sum_{n=-\infty}^{\infty}f_{n}^{\n},\nonumber\\
B^{{\rm inc}}_{lm\omega}
&=&\omega^{-1}\left[K_{\n}-ie^{-i\pi\n}
\frac{\sin\pi(\n-s+i\e)}{\sin\pi(\n+s-i\e)}K_{-\n-1}\right]
A_{+}^{\n}e^{-i\e\ln\e},\nonumber \\
B^{{\rm ref}}_{lm\omega}
&=&\omega^{-1-2s}[K_{\n}+ie^{i\pi\n}K_{-\n-1}]A_{-}^
{\n}e^{i\e\ln\e},
\label{eq:asymp_amp}
\end{eqnarray}
where
\begin{eqnarray}
A_{+}^{\n}&=&2^{-1+s-i\e}e^{-\frac{\pi\e}{2}}e^{\frac{\pi}{2}i(\n+1-s)}
\frac{\G(\n+1-s+i\e)}{\G(\n+1+s-i\e)}\sum_{n=-\infty}^{+\infty}f_{n}^{\n},
\nonumber \\
A_{-}^{\n}&=&2^{-1-s+i\e}e^{-\frac{\pi\e}{2}}e^{\frac{-\pi}{2}i(\n+1+s)}
\sum_{n=-\infty}^{+\infty}(-1)^{n}\frac{(\n+1+s-i\e)_{n}}{(\n+1-s+i\e)_{n}}
f_{n}^{\n}.
\end{eqnarray}

We can also obtain an analytical expression for $R_{lm\omega}^{{\rm up}}$ by 
transforming $R_{\rm C}$. We note that an analytical property of the 
confluent hypergeometric function,
\begin{eqnarray}
\Phi(\a,\c,x)=\frac{\G(\c)}{\G(\c-\a)}e^{i\a\pi\sigma}\Psi(\a,\c,x)+
\frac{\G(\c)}{\G(\z)}e^{i\pi(\a-\c)\sigma}\Psi(\c-\a,\c,-x).
\end{eqnarray}
Here $\sigma={\rm sgn}[{\rm Im}(x)]$ and $\Psi(a,c,x)$ is the irregular
confluent hypergeometric function. 
We then rewrite $R_{{\rm C}}^{\n}$ as
\begin{eqnarray}
R_{{\rm C}}^{\n}=R_{+}^{\n}+R_{-}^{\n},
\end{eqnarray}
where
\begin{eqnarray}
R_{+}^{\n}&=& 2^{\n}e^{-\pi \e}e^{i\pi(\n+1-s)}
\frac{\G(\n+1-s+i\e)}{\G(\n+1+s-i\e)}
e^{-iz}z^{\n+i(\e+\t)/2}(z-\e\kappa)^{-s-i(\e+\t)/2}\nonumber\\
&\times&  \sum_{n=-\infty}^{\infty}i^n
f_n^{\n}(2z)^n\Psi(n+\n+1-s+i\e,2n+2\n+2;2iz),\\
&&\nonumber\\
R_{-}^{\n}&=& 2^{\n}e^{-\pi \e}e^{-i\pi(\n+1+s)}
e^{iz}z^{\n+i(\e+\t)/2}(z-\e\kappa)^{-s-i(\e+\t)/2}\sum_{n=-\infty}^{\infty}i^n\nonumber\\
&\times&  \frac{(\n+1+s-i\e)_n}{(\n+1-s+i\e)_n}
f_n^{\n}(2z)^n\Psi(n+\n+1+s-i\e,2n+2\n+2;-2iz).
\end{eqnarray}

For large $\mid x\mid$, $\Psi(\a,\c,x)$ behaves as
\begin{eqnarray}
\Psi(\a,\b,x)\rightarrow x^{-\a}\, {\rm as}\, \mid x\mid\rightarrow \infty.  
\end{eqnarray}
We therefore have 
\begin{eqnarray}
R_{+}^{\n}&=&A_{+}^{\n}z^{-1}e^{-i(z+\e\ln z)},\nonumber \\
R_{-}^{\n}&=&A_{-}^{\n}z^{-1-2s}e^{i(z+\e\ln z)}.
\end{eqnarray}
We can see that the functions $R_{+}^{\n}$ and $R_{-}^{\n}$ are ingoing-wave 
and outgoing-wave solutions at infinity, respectively. In particular, we have 
the upgoing solution expressed in terms of Coulomb wave functions as 
\begin{eqnarray}
\label{eq:RuptoR-}
R_{lm\omega}^{{\rm up}}=R_{-}^{\n}.
\end{eqnarray}

Now, we can obtain an analytical expression of the asymptotic amplitude, 
$C_{lm\omega}^{{\rm trans}}$, of $R_{lm\omega}^{{\rm up}}$ defined in Eq. (\ref{eq:Rup-asymp}). 
We find
\begin{eqnarray}
C^{{\rm trans}}_{lm\omega}=\omega^{-1-2s}e^{i\e\ln\e}A_{-}^{\n}.
\label{eq:Ctrans}
\end{eqnarray}

\section{Numerical methods}
\label{sec:NM}

In order to compute the homogeneous solutions of the Teukolsky equation,
we first compute the eigenvalue $\lambda$ of the spin weighted spheroidal harmonics. This is discussed in Appendix \ref{sec:SWSH}. 
In this section, we discuss a numerical method to compute the homogeneous
solution of the radial part of the Teukolsky equation, assuming
$\lambda$ is given. 
The eigenvalue $\lambda$ can be computed similar in 
to $\nu$. 
Although the MST formalism was developed for arbitrary values of the spin $s$, 
in the rest of paper we consider only the case $s=-2$, because 
this is important to evaluate gravitational waves. 

\subsection{Continued fractions}

Because the computation of continued fractions is very important in our
numerical method, we first review {\it Steed's algorithm} to compute
continued fractions \cite{BFSG}. 
Let us define a continued fraction by
\begin{eqnarray}
\label{eq:program hn}
h_{n}\equiv\frac{A_{n}}{B_{n}}
=b_{0}
+\frac{a_{1}}{b_{1}+}\frac{a_{2}}{b_{2}+}\frac{a_{3}}{b_{3}+}\cdots\frac{a_{n}}{b_{n}}.
\end{eqnarray}
It can be shown that 
$A_{n}$ and $B_{n}$ satisfy the recurrence relations
\begin{eqnarray}
\left(
\begin{array}{c}
A_{n}\\
B_{n}\\
\end{array}
\right)=\left(
	\begin{array}{cc}
	A_{n-1}&A_{n-2}\\
	B_{n-1}&B_{n-2}\\
	\end{array}
	\right)\left(
	\begin{array}{c}
	b_{n}\\
	a_{n}\\
	\end{array}
	\right),
\ \ 
\left(
\begin{array}{cc}
A_{0}&A_{-1}\\
B_{0}&B_{-1}\\
\end{array}
\right)=\left(
\begin{array}{cc}
b_{0}&1\\
1&0\\
\end{array}
\right).
\end{eqnarray}
Here, we introduce the quantity $D_{n}=B_{n-1}/B_{n}$. Then, the above
recurrence relation of $B_{n}$ is identical to
\begin{eqnarray}
\label{eq:program id}
D_{n}=\frac{1}{b_{n}+a_{n}D_{n-1}}.
\end{eqnarray}
We can rewrite the right-hand side of Eq. (\ref{eq:program hn}) in
terms of $D_{n}$ as 
\begin{eqnarray}
h_{n}=\frac{A_{n}}{B_{n}}
&=&\frac{b_{n}A_{n-1}+a_{n}A_{n-2}}{b_{n}B_{n-1}+a_{n}B_{n-2}},\nonumber \\
&=&\frac{b_{n}(A_{n-1}/B_{n-1})+a_{n}(A_{n-2}B_{n-2}/B_{n-2}B_{n-1})}
{b_{n}+a_{n}(B_{n-2}/B_{n-1})},\nonumber \\
&=&(h_{n-1}b_{n}+h_{n-2}D_{n-1}a_{n})D_{n}. 
\end{eqnarray}
Furthermore, from Eq. (\ref{eq:program id}), 
we have the relation $a_{n}D_{n-1}D_{n}=1-b_{n}D_{n}$,
which then yields the difference between $h_{n}$ and $h_{n-1}$,
\begin{eqnarray}
\label{eq:Steedhn}
\Delta h_{n}&\equiv&h_{n}-h_{n-1},\nonumber \\
&=&(b_{n}D_{n}-1)(h_{n-1}-h_{n-2}),\nonumber \\
&=&(b_{n}D_{n}-1)\Delta h_{n-1}.
\end{eqnarray}
{\it Steed's algorithm} is summarized as follows. 
We first set $h_{0}=b_{0}$, $D_{1}=1/b_{1}$, $\Delta h_{1}=a_{1}/b_{1}$ and
$h_{1}=h_{0}+\Delta h_{1}$.
We then compute $h_{n}$ beginning with $n=2$ until $\mid \Delta h_{n}/h_{n}\mid$ is 
smaller than the required accuracy. 

\subsection{Determination of the renormalized angular momentum $\nu$}
\label{sec:about nu}

The parameter $\nu$ is determined as a solution of 
Eq. (\ref{eq:determine_nu}). 
In particular, we set $n=0$ and obtain 
\begin{equation}
g(\nu)\equiv \beta_0^\nu+\alpha_0^\nu \tilde{R}_{1}+\gamma_0^\nu \tilde{L}_{-1}=0,
\label{eq:gnu}
\end{equation}
where $\tilde{R}_{1}$ and $\tilde{L}_{-1}$ are expressed by the continued
fractions Eqs. (\ref{eq:tildeRn}) and (\ref{eq:tildeLn}), respectively. 


\begin{figure}[htbp]
\begin{center}
\resizebox{10.0cm}{!}{\includegraphics{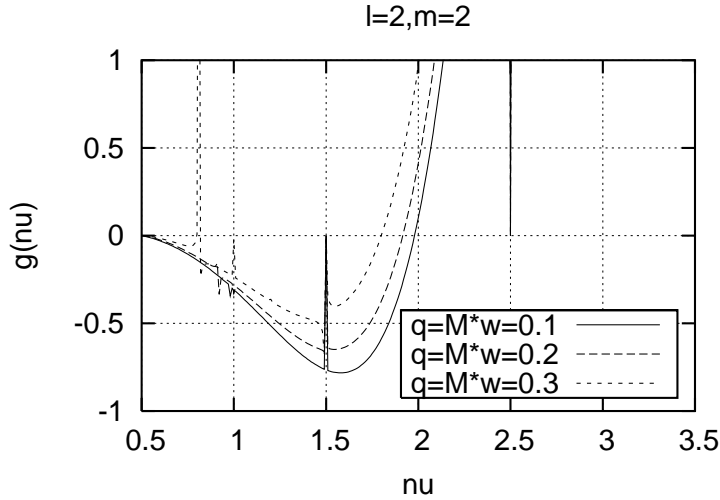}}
\caption{$g(\nu)$ for $\ell=2, m=2$.}
\label{fig:fnu22}
\end{center}
\end{figure}

\begin{figure}[htbp]
\begin{center}
\resizebox{10.0cm}{!}{\includegraphics{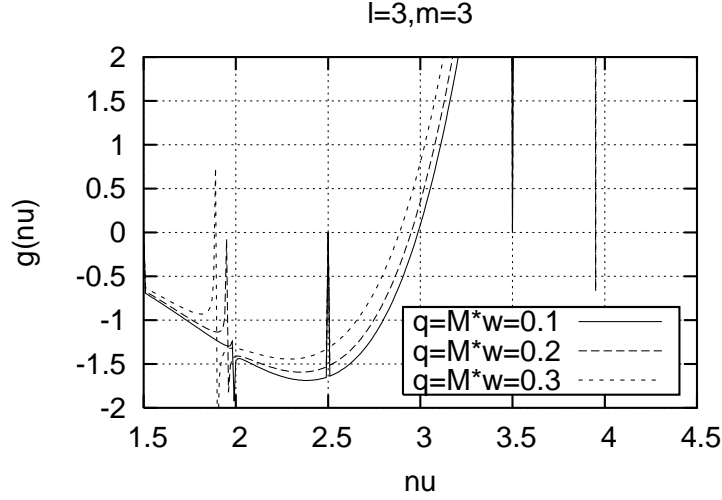}}
\caption{$g(\nu)$ for $\ell=3, m=3$.}
\label{fig:fnu33}
\end{center}
\end{figure}

\begin{figure}[htbp]
\begin{center}
\resizebox{10.0cm}{!}{\includegraphics{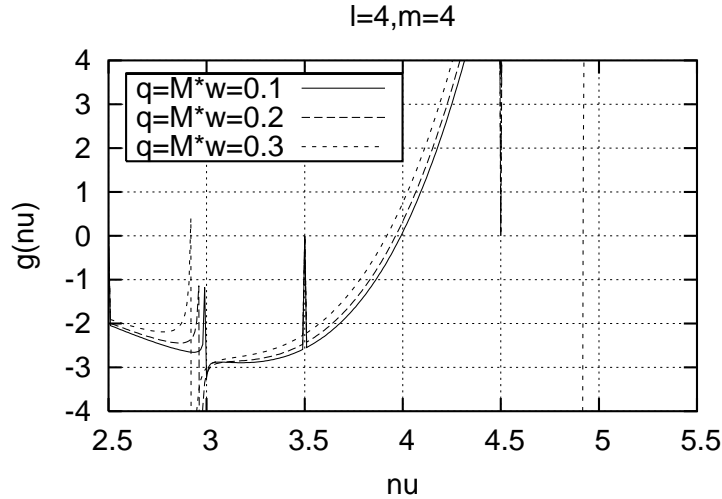}}
\caption{$g(\nu)$ for $\ell=4, m=4$.}
\label{fig:fnu44}
\end{center}
\end{figure}

\begin{figure}[htbp]
\begin{center}
\resizebox{10.0cm}{!}{\includegraphics{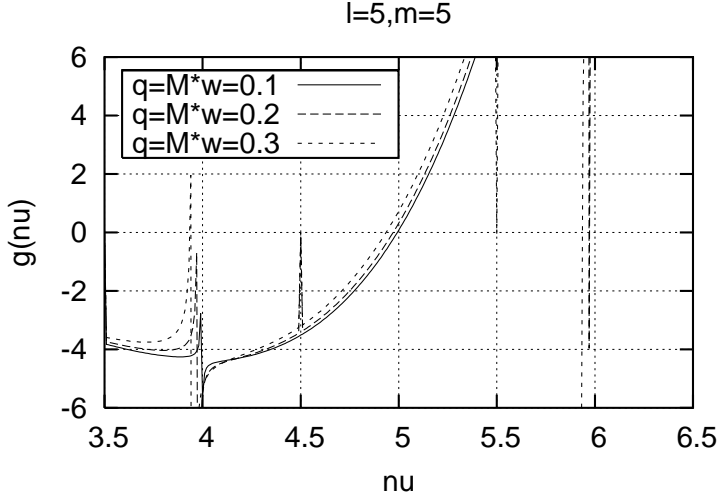}}
\caption{$g(\nu)$ for $\ell=5, m=5$.}
\label{fig:fnu55}
\end{center}
\end{figure}

{\tiny
\begin{table}[htbp]
\caption{Maximum values of $M\omega$ for which real $\nu$ is found.}
\begin{center}
\begin{tabular}{cc|ccc||cc|ccc}
\hline \hline 
$\ell$&$m$&$q=0.9$&$q=0$&$q=-0.9$&
$\ell$&$m$&$q=0.9$&$q=0$&$q=-0.9$
\\ \hline 
2 & 2  &0.39 &0.36 & 0.32 & 8 & 8  &1.08 &1.00 & 0.92 \\
2 & 1  &0.44 &0.36 & 0.34 & 8 & 7  &1.07 &1.00 & 0.93 \\
2 & 0  &0.38 &0.36 & 0.38 & 8 & 6  &1.06 &1.00 & 0.94 \\
3 & 3  &0.58 &0.53 & 0.45 & 8 & 5  &1.05 &1.00 & 0.95 \\
3 & 2  &0.61 &0.53 & 0.48 & 8 & 4  &1.04 &1.00 & 0.96 \\
3 & 1  &0.59 &0.53 & 0.51 & 8 & 3  &1.03 &1.00 & 0.97 \\
3 & 0  &0.55 &0.53 & 0.55 & 8 & 2  &1.02 &1.00 & 0.98 \\
4 & 4  &0.73 &0.66 & 0.57 & 8 & 1  &1.01 &1.00 & 0.99 \\
4 & 3  &0.73 &0.66 & 0.60 & 8 & 0  &1.00 &1.00 & 1.00 \\
4 & 2  &0.72 &0.66 & 0.63 & 9 & 9  &1.14 &1.06 & 0.99 \\
4 & 1  &0.70 &0.66 & 0.65 & 9 & 8  &1.13 &1.06 & 1.00 \\
4 & 0  &0.68 &0.66 & 0.68 & 9 & 7  &1.12 &1.06 & 1.00 \\
5 & 5  &0.84 &0.77 & 0.68 & 9 & 6  &1.12 &1.06 & 1.01 \\
5 & 4  &0.83 &0.77 & 0.70 & 9 & 5  &1.11 &1.06 & 1.02 \\
5 & 3  &0.82 &0.77 & 0.72 & 9 & 4  &1.10 &1.06 & 1.03 \\
5 & 2  &0.80 &0.77 & 0.74 & 9 & 3  &1.09 &1.06 & 1.04 \\
5 & 1  &0.79 &0.77 & 0.76 & 9 & 2  &1.08 &1.06 & 1.05 \\
5 & 0  &0.77 &0.77 & 0.77 & 9 & 1  &1.07 &1.06 & 1.06 \\
6 & 6  &0.93 &0.85 & 0.77 & 9 & 0  &1.06 &1.06 & 1.06 \\
6 & 5  &0.92 &0.85 & 0.79 & 10& 10 &1.20 &1.12 & 1.05 \\
6 & 4  &0.91 &0.85 & 0.80 & 10& 9  &1.19 &1.12 & 1.05 \\
6 & 3  &0.89 &0.85 & 0.82 & 10& 8  &1.19 &1.12 & 1.06 \\
6 & 2  &0.88 &0.85 & 0.83 & 10& 7  &1.18 &1.12 & 1.07 \\
6 & 1  &0.87 &0.85 & 0.84 & 10& 6  &1.17 &1.12 & 1.08 \\
6 & 0  &0.86 &0.85 & 0.86 & 10& 5  &1.16 &1.12 & 1.09 \\
7 & 7  &1.01 &0.93 & 0.85 & 10& 4  &1.16 &1.12 & 1.09 \\
7 & 6  &1.00 &0.93 & 0.86 & 10& 3  &1.15 &1.12 & 1.10 \\
7 & 5  &0.99 &0.93 & 0.88 & 10& 2  &1.14 &1.12 & 1.11 \\
7 & 4  &0.98 &0.93 & 0.89 & 10& 1  &1.13 &1.12 & 1.12 \\
7 & 3  &0.96 &0.93 & 0.90 & 10& 0  &1.12 &1.12 & 1.12 \\
7 & 2  &0.95 &0.93 & 0.91 &   &    &     &     &      \\
7 & 1  &0.94 &0.93 & 0.92 &   &    &     &     &      \\
7 & 0  &0.93 &0.93 & 0.93 &   &    &     &     &      \\
\hline \hline
\end{tabular}
\end{center}
\label{tab:real-nu-max}
\end{table}
}

In order to search for a root of 
the implicit equation $g(\nu)=0$, we can use various numerical techniques.
Among them, we adopt {\it Brent's algorithm} 
(e.g. Ref.~\citen{Recipes}). 
In order to search for a root of implicit equations,
we need an initial value of the search which is not very far away 
from the root of the equation.
In the case that $M\omega$ is small, 
we can use an analytic expression of $\nu$ as the initial value.
Among the infinite number of roots, there is an analytic expression
of $\nu$ in the form of a series of $\epsilon\equiv 2M\omega$ given by 
\begin{eqnarray}
\label{eq:nusol}
\nu&=& \ell+{1\over{2\ell+1}}\left[-2-{s^2\over{\ell(\ell+1)}}
+{[(\ell+1)^2-s^2]^2
\over{(2\ell+1)(2\ell+2)(2\ell+3)}}
\right.
\nonumber\\
&&
\left.
-{(\ell^2-s^2)^2\over{(2\ell-1)2\ell(2\ell+1)}}\right]
\epsilon^2+O(\epsilon^3). 
\end{eqnarray}

When $M\omega$ is not very large 
(less than $\sim 0.36$ in the case $q=0$), 
this expression is useful as an initial value 
for the root search algorithm. 
Note that the analytic expression for $\nu$ truncated at $O(\epsilon^2)$ is 
always slightly less than $\ell$. Thus, in practice, it is sufficient
to search for a root in the range $\ell-1/2<\nu<\ell$. 
We can see in Figs. \ref{fig:fnu22}--\ref{fig:fnu55} that
the function $g(\nu)$ is smooth in this range.
Although numerically we find poles at half integer values of $\nu$, 
it is not difficult to find a root in this region. 
In fact, we can find a root as long as $M\omega\lesssim 0.36$ 
in the case $\ell=m=2$ and $q=0$. 

However, the situation is different when $M\omega$ becomes larger. 
In this case, 
$\nu$ approaches $\ell-1/2$, and beyond a certain value of $M\omega$, 
it becomes impossible to find roots $\nu$. 
The maximum values of $M\omega$ for which real $\nu$ can be found 
in the region $\ell-1/2<\nu<\ell$ are 
listed in Table \ref{tab:real-nu-max}. The maximum values depend on 
$\ell,m$ and $q$. This maximum value of $M\omega$ increases with $\ell$. 
The maximum value is larger for $q>0$ than for $q<0$. 
The behavior of $\nu$ beyond $M\omega$ in Table \ref{tab:real-nu-max} depends 
on $\ell$ and $m$. 
In some cases, we can find $\nu$ as a real value. However, in other cases, 
there is no root of Eq. (\ref{eq:gnu}) 
for any real value of $\nu$. 

\begin{figure}
\begin{center}
\resizebox{10.0cm}{!}{\includegraphics{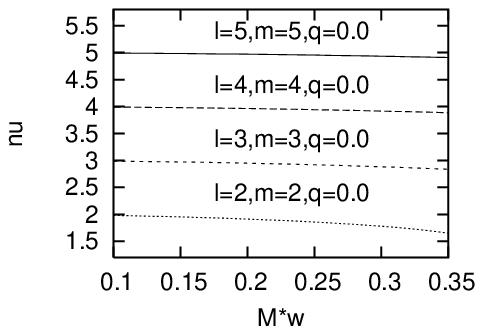}}
\resizebox{10.0cm}{!}{\includegraphics{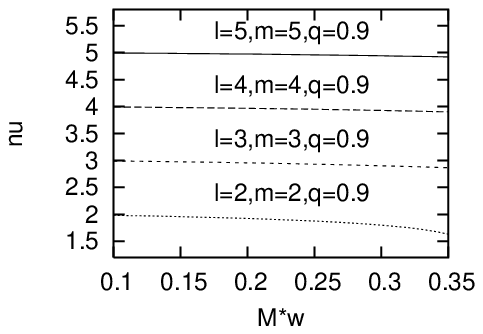}}
\resizebox{10.0cm}{!}{\includegraphics{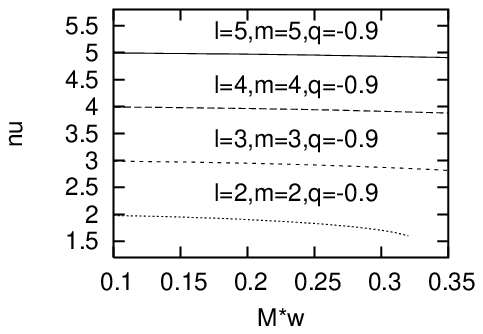}}
\caption{$\nu$ as a function of $M\omega$ for 
$\ell=m=2$ to $\ell=m=5$ for $q=0$, $0.9$ and $-0.9$. }
\label{fig:plotnu}
\end{center}
\end{figure}

We note that we have attempted to find a complex root $\nu$. 
Our preliminary investigation suggests that there is such a root 
in the complex value of $\nu$ when we can not find a real root. 
However, more work is 
needed to confirm this, and we leave this as a future project. 
In the rest of this paper, we concentrate on the case of real $\nu$ 
and the region of $\epsilon$ in which we can find such a root. 
The values of $\nu$ for various values of $M\omega=\epsilon/2$, 
$\ell$, and $q$ are plotted in Fig. \ref{fig:plotnu}. 

Note that there are some roots at integer and half integer values of $\nu$
that are independent of $\ell,m,q$, and $\omega$. 
Although they are solutions of $g(\nu)=0$, 
they do not connect the two minimal solutions in the limits $n\rightarrow \pm\infty$. 

When $\nu$ is a root of Eq. (\ref{eq:gnu}), 
the quantities $\nu+k$ ($k=\pm 1, \pm 2, \cdots$) are also roots of
Eq. (\ref{eq:gnu}), 
since $\nu$ only appears as $\nu+n$ in the continued fractions
of Eq. (\ref{eq:gnu}). This fact can be checked numerically for $k=\pm 1$. 
However, as we can see in Figs. (\ref{fig:fnu22})--(\ref{fig:fnu55}), 
the function $g(\nu)$ around the roots $\nu\pm 1$ is a very steep 
function. 
When $|k|$ is larger than $2$, the slope of $g(\nu)$ becomes much steeper, 
and it becomes difficult to find the root $\nu+k$. 

\begin{table}[htbp]
\begin{center}
\caption{$\nu$ for $M\omega=0.1$.}
\begin{tabular}{cr|ccc}
\hline \hline 
$\ell$&$m$&$q=-0.9$&$q=0$&$q=0.9$\\ \hline 
$2$&$2$&$1.9780030721$&$1.9793154547$&$1.9805149449$\\
$2$&$1$&$1.9787154140$&$1.9793154547$&$1.9799783120$\\
$2$&$0$&$1.9793759881$&$1.9793154547$&$1.9793759881$\\
$2$&$-1$&$1.9799783120$&$1.9793154547$&$1.9787154140$\\
$2$&$-2$&$1.9805149449$&$1.9793154547$&$1.9780030721$\\
$3$&$3$&$2.9871135866$&$2.9875539197$&$2.9879712028$\\
$3$&$2$&$2.9872683427$&$2.9875539197$&$2.9878392372$\\
$3$&$1$&$2.9874180082$&$2.9875539197$&$2.9877032005$\\
$3$&$0$&$2.9875628720$&$2.9875539197$&$2.9875628720$\\
$3$&$-1$&$2.9877032005$&$2.9875539197$&$2.9874180082$\\
$3$&$-2$&$2.9878392372$&$2.9875539197$&$2.9872683427$\\
$3$&$-3$&$2.9879712028$&$2.9875539197$&$2.9871135866$\\
$4$&$4$&$3.9906679155$&$3.9909066870$&$3.9911355283$\\
$4$&$3$&$3.9907299644$&$3.9909066870$&$3.9910804817$\\
$4$&$2$&$3.9907909277$&$3.9909066870$&$3.9910245146$\\
$4$&$1$&$3.9908508359$&$3.9909066870$&$3.9909676019$\\
$4$&$0$&$3.9909097180$&$3.9909066870$&$3.9909097180$\\
$4$&$-1$&$3.9909676019$&$3.9909066870$&$3.9908508359$\\
$4$&$-2$&$3.9910245146$&$3.9909066870$&$3.9907909277$\\
$4$&$-3$&$3.9910804817$&$3.9909066870$&$3.9907299644$\\
$4$&$-4$&$3.9911355283$&$3.9909066870$&$3.9906679155$\\
$5$&$5$&$4.9926264974$&$4.9927797435$&$4.9929276871$\\
$5$&$4$&$4.9926581101$&$4.9927797435$&$4.9928989965$\\
$5$&$3$&$4.9926893758$&$4.9927797435$&$4.9928700024$\\
$5$&$2$&$4.9927203003$&$4.9927797435$&$4.9928406998$\\
$5$&$1$&$4.9927508893$&$4.9927797435$&$4.9928110837$\\
$5$&$0$&$4.9927811486$&$4.9927797435$&$4.9927811486$\\
$5$&$-1$&$4.9928110837$&$4.9927797435$&$4.9927508893$\\
$5$&$-2$&$4.9928406998$&$4.9927797435$&$4.9927203003$\\
$5$&$-3$&$4.9928700024$&$4.9927797435$&$4.9926893758$\\
$5$&$-4$&$4.9928989965$&$4.9927797435$&$4.9926581101$\\
$5$&$-5$&$4.9929276871$&$4.9927797435$&$4.9926264974$\\
$6$&$6$&$5.9938843772$&$5.9939919474$&$5.9940963285$\\
$6$&$5$&$5.9939027772$&$5.9939919474$&$5.9940793756$\\
$6$&$4$&$5.9939210379$&$5.9939919474$&$5.9940622984$\\
$6$&$3$&$5.9939391609$&$5.9939919474$&$5.9940450957$\\
$6$&$2$&$5.9939571478$&$5.9939919474$&$5.9940277660$\\
$6$&$1$&$5.9939750003$&$5.9939919474$&$5.9940103079$\\
$6$&$0$&$5.9939927198$&$5.9939919474$&$5.9939927198$\\
$6$&$-1$&$5.9940103079$&$5.9939919474$&$5.9939750003$\\
$6$&$-2$&$5.9940277660$&$5.9939919474$&$5.9939571478$\\
$6$&$-3$&$5.9940450957$&$5.9939919474$&$5.9939391609$\\
$6$&$-4$&$5.9940622984$&$5.9939919474$&$5.9939210379$\\
$6$&$-5$&$5.9940793756$&$5.9939919474$&$5.9939027772$\\
$6$&$-6$&$5.9940963285$&$5.9939919474$&$5.9938843772$\\
\hline \hline
\end{tabular}
\end{center}
\label{tab:nu}
\end{table}

\subsection{Expansion coefficients}

Once we have $\nu$, 
it is straightforward to evaluate the expansion coefficients
using the three-term recurrence relations Eq. (\ref{eq:3term}). 
As discussed in $\S$ \ref{sec:MST}, the expansion coefficients
$\{a_n\}$ must be the minimal solution $\{f_n\}$. 
It is well known that 
the minimal solution cannot be computed numerically with forward 
recursion of the three-term recurrence relations from 
$n=0$ to $\pm\infty$. This is because a small numerical errors in
the minimal solution $f_n$ 
contain dominant parts of the solution of the recurrence relation.
For this reason, the numerical solution obtained using forward recursion will be dominated 
by errors after several recursions, since the dominant solutions 
grow very rapidly. 

In such a situation, we can use the continued fractions given in 
Eqs. (\ref{eq:tildeRn}) and (\ref{eq:tildeLn}) to evaluate the minimal solution. 
First, we set the initial value of $f_n$ as $f_0=1$. 
We evaluate $f_1$ from $f_1/f_0$, which is evaluated by 
the continued fraction, the right-hand side of Eq. (\ref{eq:tildeRn}) with $n=1$. 
We then obtain $f_1$ as $f_1=\tilde{R}_1 f_0$. 
In the same way, we can evaluate $f_n$ $(n>1)$ from $f_{n-1}$ and $\tilde{R}_n$ 
recursively. 
For $n<0$,  we use the same algorithm and evaluate $f_{n}$ 
from $f_{n+1}$ and $\tilde{L}_n$. 

\subsection{Homogeneous solutions}

Given the eigenvalue of the spin weighted spheroidal harmonics $\lambda$, 
the renormalized angular momentum $\nu$,
and the expansion coefficients ${f_n}$, it is straightforward to compute the 
asymptotic amplitude of the homogeneous solutions, Eqs. (\ref{eq:asymp_amp})
and (\ref{eq:Ctrans}).
It is also straightforward to compute the 
homogeneous functions using Eq. (\ref{eq:Rin}) or Eq. (\ref{eq:RuptoR-}). 
We found that the convergence of the series of 
hypergeometric functions or Coulomb wave functions and 
that of the formulas for the asymptotic amplitude are very rapid. 
This is because the expansion coefficients $\{f_n^\nu\}$,
which constitute a minimal solution of the three-term recurrence relation,
decrease very rapidly as $|n|$ becomes large,
as can be seen from Eq. (\ref{eq:asymptotics of minimal solution}).
For example, we only need to carry out the summation in Eq. (\ref{eq:series of Rin})
from $n=0$ to $n\sim\pm 20$ to have an accuracy $10^{-16}$ 
in the case $\e<0.4$. 

The hypergeometric function $F(\alpha,\beta;\gamma,x)$ is computed using 
the transformation formula
\begin{eqnarray}
F(\alpha,\beta;\gamma,x)&=&(1-x)^{-\alpha}
\frac{\Gamma(\gamma)\Gamma(\beta-\alpha)}{\Gamma(\beta)\Gamma(\gamma-\alpha)}
F(\alpha,\gamma-\beta;\alpha-\beta+1;\frac{1}{1-x})
\nonumber\\
&&\quad 
+(1-x)^{-\beta}
\frac{\Gamma(\gamma)\Gamma(\alpha-\beta)}{\Gamma(\alpha)\Gamma(\gamma-\beta)}
F(\beta,\gamma-\alpha;\beta-\alpha+1;\frac{1}{1-x}),
\label{eq:hypertrans}
\end{eqnarray}
and using the Gauss hypergeometric series for the hypergeometric
functions on the right-hand side of Eq. (\ref{eq:hypertrans}).
For the gamma function, 
we use the routines available on the world wide web\cite{gammafunction}.

\subsection{Gravitational wave luminosity}
\label{sec:luminosity}

In order to check the accuracy of our numerical code, 
we calculated the gravitational wave flux from a point particle in circular
orbits on the equatorial plane around a Kerr black hole. 
The formula for the luminosity is given in Appendix \ref{sec:Teukolsky}. 

Our computations presented here were done using a double precision code. 
In Tables \ref{tab:energy_flux6}--\ref{tab:energy_flux1000}, 
we list the luminosities for $r_0=6$, $10$, $100$ and $1000$ 
from $\ell=2$ to $6$ and for $q=0$ and $q=\pm 0.9$. 
We also list the total luminosity 
for various values $q$ and $r_0$ in Table \ref{tab:energy_flux}. 

We compared these results with those of 
Tagoshi and Nakamura\cite{TN} in the Schwarzschild case $q=0$.
The results are given in Table \ref{tab:flux-compare_TN} for cases $r_0=10M$.
We find that our results agree with them with relative error $10^{-14}$ -- $10^{-15}$. 
Because the estimated accuracy of Ref. \citen{TN} is more than 20 significant figures, 
we estimate that the accuracy of our code is about 13--14 significant figures. 
We also compare our results with those given by Kennefick \cite{GK2}
in the Kerr case $q\neq 0$. 
The results are given in Table \ref{tab:flux-compare_kenn} for cases $r_0=3M$ and $q=0.998$. 
We find that 
our results agree with them with relative error $10^{-6}$ -- $10^{-7}$. 
This is consistent with the estimated accuracy in his works. 
Although we do not have a quantitative estimate of the accuracy for the luminosity
in the Kerr case, 
because $\nu$ is computed with accuracy in the Schwarzschild case, 
we expect that the accuracy of our results is also about 13--14 significant figures in the Kerr case. 

\begin{table}[htbp]
\begin{center}
\caption{The gravitational wave luminosity for $r_0=6M$.}
\begin{tabular}{cc|ccc}
\hline \hline 
$\ell$&$\mid m\mid$&$q=-0.9$&$q=0$&$q=0.9$\\ \hline 
$2$&$2$&$1.4672643416\times 10^{-3}$&$7.3475638881\times 10^{-4}$&$4.6183912921\times 10^{-4}$\\
$2$&$1$&$2.4135187915\times 10^{-5}$&$5.0413451839\times 10^{-6}$&$6.6947435866\times 10^{-7}$\\
$3$&$3$&$3.4798321412\times 10^{-4}$&$1.4534938751\times 10^{-4}$&$8.0343009373\times 10^{-5}$\\
$3$&$2$&$1.0938697387\times 10^{-5}$&$2.0567575315\times 10^{-6}$&$2.9212522769\times 10^{-7}$\\
$3$&$1$&$4.3891293595\times 10^{-8}$&$1.1384270937\times 10^{-8}$&$4.3010746650\times 10^{-9}$\\
$4$&$4$&$1.0477344711\times 10^{-4}$&$3.5849943696\times 10^{-5}$&$1.7293785265\times 10^{-5}$\\
$4$&$3$&$3.9195590828\times 10^{-6}$&$6.2995356560\times 10^{-7}$&$8.6369900044\times 10^{-8}$\\
$4$&$2$&$7.5231697792\times 10^{-8}$&$1.7547382460\times 10^{-8}$&$6.1994559709\times 10^{-9}$\\
$4$&$1$&$4.5312182607\times 10^{-11}$&$7.0534448975\times 10^{-12}$&$9.3343722027\times 10^{-13}$\\
$5$&$5$&$3.4447343626\times 10^{-5}$&$9.5758155537\times 10^{-6}$&$4.0148159555\times 10^{-6}$\\
$5$&$4$&$1.3473368296\times 10^{-6}$&$1.8048918361\times 10^{-7}$&$2.2863071277\times 10^{-8}$\\
$5$&$3$&$5.0155535469\times 10^{-8}$&$1.0218507481\times 10^{-8}$&$3.3101913526\times 10^{-9}$\\
$5$&$2$&$2.3873981915\times 10^{-10}$&$3.1974516361\times 10^{-11}$&$3.9575575812\times 10^{-12}$\\
$5$&$1$&$4.8597838388\times 10^{-14}$&$9.3414524994\times 10^{-15}$&$2.8261985682\times 10^{-15}$\\
$6$&$6$&$1.1820336377\times 10^{-5}$&$2.6570852921\times 10^{-6}$&$9.6579842085\times 10^{-7}$\\
$6$&$5$&$4.5987487190\times 10^{-7}$&$5.0650467610\times 10^{-8}$&$5.7941986759\times 10^{-9}$\\
$6$&$4$&$2.5069319096\times 10^{-8}$&$4.3708099159\times 10^{-9}$&$1.2798607028\times 10^{-9}$\\
$6$&$3$&$2.7050690849\times 10^{-10}$&$3.0671520252\times 10^{-11}$&$3.4745761467\times 10^{-12}$\\
$6$&$2$&$8.1558279125\times 10^{-13}$&$1.3490851660\times 10^{-13}$&$3.6798960699\times 10^{-14}$\\
$6$&$1$&$2.8413832649\times 10^{-17}$&$3.3772691941\times 10^{-18}$&$3.9283478383\times 10^{-19}$\\
\hline \hline
\end{tabular}
\label{tab:energy_flux6}
\end{center}
\end{table}

\begin{table}[htbp]
\begin{center}
\caption{The gravitational wave luminosity for $r_0=10M$.}
\begin{tabular}{cc|ccc}
\hline \hline 
$\ell$&$\mid m\mid$&$q=-0.9$&$q=0$&$q=0.9$\\ \hline 
$2$&$2$&$6.8120258138\times 10^{-5}$&$5.3687954791\times 10^{-5}$&$4.4546001102\times 10^{-5}$\\
$2$&$1$&$5.0656468066\times 10^{-7}$&$1.9316093512\times 10^{-7}$&$5.2735787002\times 10^{-8}$\\
$3$&$3$&$8.7890229491\times 10^{-6}$&$6.4260827562\times 10^{-6}$&$5.0413559470\times 10^{-6}$\\
$3$&$2$&$1.2628541265\times 10^{-7}$&$4.7959164616\times 10^{-8}$&$1.4271936289\times 10^{-8}$\\
$3$&$1$&$9.7967935639\times 10^{-10}$&$5.7148989126\times 10^{-10}$&$3.7294384569\times 10^{-10}$\\
$4$&$4$&$1.4106496214\times 10^{-6}$&$9.5396003949\times 10^{-7}$&$7.0641315437\times 10^{-7}$\\
$4$&$3$&$2.4037498342\times 10^{-8}$&$8.7787575252\times 10^{-9}$&$2.6507201263\times 10^{-9}$\\
$4$&$2$&$9.3303939387\times 10^{-10}$&$5.2622453090\times 10^{-10}$&$3.3420025690\times 10^{-10}$\\
$4$&$1$&$4.1524457529\times 10^{-13}$&$1.4575856423\times 10^{-13}$&$4.2218942492\times 10^{-14}$\\
$5$&$5$&$2.4401253104\times 10^{-7}$&$1.5241547646\times 10^{-7}$&$1.0642982525\times 10^{-7}$\\
$5$&$4$&$4.3181478959\times 10^{-9}$&$1.4921162749\times 10^{-9}$&$4.4382292146\times 10^{-10}$\\
$5$&$3$&$3.3947862663\times 10^{-10}$&$1.8291013252\times 10^{-10}$&$1.1221183734\times 10^{-10}$\\
$5$&$2$&$1.1385202800\times 10^{-12}$&$3.8193532372\times 10^{-13}$&$1.0943525441\times 10^{-13}$\\
$5$&$1$&$4.6491185735\times 10^{-16}$&$2.3676371874\times 10^{-16}$&$1.3782555714\times 10^{-16}$\\
$6$&$6$&$4.3683996396\times 10^{-8}$&$2.5182131568\times 10^{-8}$&$1.6571971666\times 10^{-8}$\\
$6$&$5$&$7.6349317589\times 10^{-10}$&$2.4746386947\times 10^{-10}$&$7.1425097921\times 10^{-11}$\\
$6$&$4$&$9.1394784860\times 10^{-11}$&$4.6633398847\times 10^{-11}$&$2.7471342862\times 10^{-11}$\\
$6$&$3$&$6.6822442144\times 10^{-13}$&$2.1238827476\times 10^{-13}$&$5.9371098714\times 10^{-14}$\\
$6$&$2$&$4.0965704268\times 10^{-15}$&$1.9763689535\times 10^{-15}$&$1.1021357357\times 10^{-15}$\\
$6$&$1$&$1.1358089132\times 10^{-19}$&$3.5977953599\times 10^{-20}$&$9.9623097231\times 10^{-21}$\\
\hline \hline
\end{tabular}
\label{tab:energy_flux10}
\end{center}
\end{table}

\begin{table}[htbp]
\begin{center}
\caption{The gravitational wave luminosity for $r_0=100M$.}
\begin{tabular}{cc|ccc}
\hline \hline 
$\ell$&$\mid m\mid$&$q=-0.9$&$q=0$&$q=0.9$\\ \hline 
$2$&$2$&$6.1883920768\times 10^{-10}$&$6.1534957152\times 10^{-10}$&$6.1208585286\times 10^{-10}$\\
$2$&$1$&$2.2912106158\times 10^{-13}$&$1.7672378807\times 10^{-13}$&$1.3144823366\times 10^{-13}$\\
$3$&$3$&$8.2148864045\times 10^{-12}$&$8.1530505104\times 10^{-12}$&$8.0957455570\times 10^{-12}$\\
$3$&$2$&$6.2294067784\times 10^{-15}$&$4.9252050773\times 10^{-15}$&$3.7844616487\times 10^{-15}$\\
$3$&$1$&$7.6794049338\times 10^{-16}$&$7.5711224891\times 10^{-16}$&$7.4713604631\times 10^{-16}$\\
$4$&$4$&$1.3404252509\times 10^{-13}$&$1.3278019616\times 10^{-13}$&$1.3161763785\times 10^{-13}$\\
$4$&$3$&$1.2347499275\times 10^{-16}$&$9.8769443625\times 10^{-17}$&$7.7063402323\times 10^{-17}$\\
$4$&$2$&$7.6509385330\times 10^{-17}$&$7.5369817847\times 10^{-17}$&$7.4317624281\times 10^{-17}$\\
$4$&$1$&$2.6875757595\times 10^{-21}$&$2.1456137700\times 10^{-21}$&$1.6707364028\times 10^{-21}$\\
$5$&$5$&$2.3376726712\times 10^{-15}$&$2.3112411600\times 10^{-15}$&$2.2870144702\times 10^{-15}$\\
$5$&$4$&$2.2627547216\times 10^{-18}$&$1.8214609703\times 10^{-18}$&$1.4332319114\times 10^{-18}$\\
$5$&$3$&$2.8901658332\times 10^{-18}$&$2.8441315803\times 10^{-18}$&$2.8016548481\times 10^{-18}$\\
$5$&$2$&$7.3789391681\times 10^{-22}$&$5.9290982541\times 10^{-22}$&$4.6565092272\times 10^{-22}$\\
$5$&$1$&$4.6833027135\times 10^{-24}$&$4.5980275148\times 10^{-24}$&$4.5192415969\times 10^{-24}$\\
$6$&$6$&$4.1958033380\times 10^{-17}$&$4.1404427415\times 10^{-17}$&$4.0898989226\times 10^{-17}$\\
$6$&$5$&$4.0369200351\times 10^{-20}$&$3.2613208135\times 10^{-20}$&$2.5791601982\times 10^{-20}$\\
$6$&$4$&$8.0084356788\times 10^{-20}$&$7.8709992082\times 10^{-20}$&$7.7444340187\times 10^{-20}$\\
$6$&$3$&$4.3274432473\times 10^{-23}$&$3.4903584660\times 10^{-23}$&$2.7554781747\times 10^{-23}$\\
$6$&$2$&$4.1046272249\times 10^{-24}$&$4.0245352816\times 10^{-24}$&$3.9505730568\times 10^{-24}$\\
$6$&$1$&$1.0005467982\times 10^{-29}$&$8.0635499224\times 10^{-30}$&$6.3602915489\times 10^{-30}$\\
\hline \hline
\end{tabular}
\label{tab:energy_flux100}
\end{center}
\end{table}

\begin{table}[htbp]
\begin{center}
\caption{The gravitational wave luminosity for $r_0=1000M$.}
\begin{tabular}{cc|ccc}
\hline \hline 
$\ell$&$\mid m\mid$&$q=-0.9$&$q=0$&$q=0.9$\\ \hline 
$2$&$2$&$6.3710447707\times 10^{-15}$&$6.3699449938\times 10^{-15}$&$6.3688661222\times 10^{-15}$\\
$2$&$1$&$1.9312210104\times 10^{-19}$&$1.7759713679\times 10^{-19}$&$1.6272619997\times 10^{-19}$\\
$3$&$3$&$8.6164205339\times 10^{-18}$&$8.6144354096\times 10^{-18}$&$8.6124927366\times 10^{-18}$\\
$3$&$2$&$5.4522769983\times 10^{-22}$&$5.0596103733\times 10^{-22}$&$4.6817315382\times 10^{-22}$\\
$3$&$1$&$7.8992896709\times 10^{-22}$&$7.8958536669\times 10^{-22}$&$7.8924917557\times 10^{-22}$\\
$4$&$4$&$1.4308259101\times 10^{-20}$&$1.4304135903\times 10^{-20}$&$1.4300106837\times 10^{-20}$\\
$4$&$3$&$1.1099463480\times 10^{-24}$&$1.0346766311\times 10^{-24}$&$9.6207662590\times 10^{-25}$\\
$4$&$2$&$8.0054989997\times 10^{-24}$&$8.0018141364\times 10^{-24}$&$7.9982051838\times 10^{-24}$\\
$4$&$1$&$2.4183146906\times 10^{-29}$&$2.2541754488\times 10^{-29}$&$2.0958722688\times 10^{-29}$\\
$5$&$5$&$2.5377450811\times 10^{-23}$&$2.5368671349\times 10^{-23}$&$2.5360100929\times 10^{-23}$\\
$5$&$4$&$2.0796404136\times 10^{-27}$&$1.9438447940\times 10^{-27}$&$1.8126926234\times 10^{-27}$\\
$5$&$3$&$3.0770492071\times 10^{-26}$&$3.0755338762\times 10^{-26}$&$3.0740495814\times 10^{-26}$\\
$5$&$2$&$6.7754641895\times 10^{-31}$&$6.3326624808\times 10^{-31}$&$5.9050352599\times 10^{-31}$\\
$5$&$1$&$4.9684166900\times 10^{-32}$&$4.9656062758\times 10^{-32}$&$4.9628506310\times 10^{-32}$\\
$6$&$6$&$4.6302492101\times 10^{-26}$&$4.6283797773\times 10^{-26}$&$4.6265561943\times 10^{-26}$\\
$6$&$5$&$3.7846535394\times 10^{-30}$&$3.5438278934\times 10^{-30}$&$3.3110410807\times 10^{-30}$\\
$6$&$4$&$8.6738999088\times 10^{-29}$&$8.6692973914\times 10^{-29}$&$8.6647902975\times 10^{-29}$\\
$6$&$3$&$4.0483295654\times 10^{-33}$&$3.7905188064\times 10^{-33}$&$3.5413297645\times 10^{-33}$\\
$6$&$2$&$4.4179556837\times 10^{-33}$&$4.4152753953\times 10^{-33}$&$4.4126466761\times 10^{-33}$\\
$6$&$1$&$9.4073281205\times 10^{-40}$&$8.8079986933\times 10^{-40}$&$8.2287305560\times 10^{-40}$\\
\hline \hline
\end{tabular}
\label{tab:energy_flux1000}
\end{center}
\end{table}

\begin{table}[htbp]
\begin{center}
\caption{The gravitational wave luminosity, up through $\ell=6$, 
for various $q$ and orbital radii $r_0/M$.}
\begin{tabular}{c|ccccccc}
\hline \hline 
$r_0/M$&$q=-0.9$&$q=-0.6$&$q=-0.3$&$q=0$&$q=0.3$
&$q=0.6$&$q=0.9$\\ \hline 
6&\hspace{-2mm}2.0073$\times 10^{-3}$&\hspace{-2mm}1.5089$\times 10^{-3}$&\hspace{-2mm}1.1707$\times 10^{-3}$&\hspace{-2mm}9.3619$\times 10^{-4}$&\hspace{-2mm}7.7061$\times 10^{-4}$&\hspace{-2mm}6.5182$\times 10^{-4}$&\hspace{-2mm}5.6555$\times 10^{-4}$\\
8&\hspace{-2mm}2.9124$\times 10^{-4}$&\hspace{-2mm}2.5217$\times 10^{-4}$&\hspace{-2mm}2.2097$\times 10^{-4}$&\hspace{-2mm}1.9593$\times 10^{-4}$&\hspace{-2mm}1.7577$\times 10^{-4}$&\hspace{-2mm}1.5948$\times 10^{-4}$&\hspace{-2mm}1.4631$\times 10^{-4}$\\
10&\hspace{-2mm}7.9272$\times 10^{-5}$&\hspace{-2mm}7.2377$\times 10^{-5}$&\hspace{-2mm}6.6505$\times 10^{-5}$&\hspace{-2mm}6.1499$\times 10^{-5}$&\hspace{-2mm}5.7229$\times 10^{-5}$&\hspace{-2mm}5.3588$\times 10^{-5}$&\hspace{-2mm}5.0488$\times 10^{-5}$\\
12&\hspace{-2mm}2.9095$\times 10^{-5}$&\hspace{-2mm}2.7287$\times 10^{-5}$&\hspace{-2mm}2.5693$\times 10^{-5}$&\hspace{-2mm}2.4289$\times 10^{-5}$&\hspace{-2mm}2.3052$\times 10^{-5}$&\hspace{-2mm}2.1965$\times 10^{-5}$&\hspace{-2mm}2.1010$\times 10^{-5}$\\
14&\hspace{-2mm}1.2794$\times 10^{-5}$&\hspace{-2mm}1.2190$\times 10^{-5}$&\hspace{-2mm}1.1646$\times 10^{-5}$&\hspace{-2mm}1.1157$\times 10^{-5}$&\hspace{-2mm}1.0717$\times 10^{-5}$&\hspace{-2mm}1.0323$\times 10^{-5}$&\hspace{-2mm}9.9699$\times 10^{-6}$\\
16&\hspace{-2mm}6.3614$\times 10^{-6}$&\hspace{-2mm}6.1235$\times 10^{-6}$&\hspace{-2mm}5.9062$\times 10^{-6}$&\hspace{-2mm}5.7079$\times 10^{-6}$&\hspace{-2mm}5.5272$\times 10^{-6}$&\hspace{-2mm}5.3627$\times 10^{-6}$&\hspace{-2mm}5.2134$\times 10^{-6}$\\
18&\hspace{-2mm}3.4599$\times 10^{-6}$&\hspace{-2mm}3.3541$\times 10^{-6}$&\hspace{-2mm}3.2565$\times 10^{-6}$&\hspace{-2mm}3.1666$\times 10^{-6}$&\hspace{-2mm}3.0837$\times 10^{-6}$&\hspace{-2mm}3.0075$\times 10^{-6}$&\hspace{-2mm}2.9376$\times 10^{-6}$\\
20&\hspace{-2mm}2.0155$\times 10^{-6}$&\hspace{-2mm}1.9640$\times 10^{-6}$&\hspace{-2mm}1.9160$\times 10^{-6}$&\hspace{-2mm}1.8714$\times 10^{-6}$&\hspace{-2mm}1.8301$\times 10^{-6}$&\hspace{-2mm}1.7918$\times 10^{-6}$&\hspace{-2mm}1.7563$\times 10^{-6}$\\
22&\hspace{-2mm}1.2397$\times 10^{-6}$&\hspace{-2mm}1.2127$\times 10^{-6}$&\hspace{-2mm}1.1874$\times 10^{-6}$&\hspace{-2mm}1.1637$\times 10^{-6}$&\hspace{-2mm}1.1416$\times 10^{-6}$&\hspace{-2mm}1.1210$\times 10^{-6}$&\hspace{-2mm}1.1018$\times 10^{-6}$\\
24&\hspace{-2mm}7.9701$\times 10^{-7}$&\hspace{-2mm}7.8197$\times 10^{-7}$&\hspace{-2mm}7.6782$\times 10^{-7}$&\hspace{-2mm}7.5452$\times 10^{-7}$&\hspace{-2mm}7.4204$\times 10^{-7}$&\hspace{-2mm}7.3033$\times 10^{-7}$&\hspace{-2mm}7.1938$\times 10^{-7}$\\
26&\hspace{-2mm}5.3155$\times 10^{-7}$&\hspace{-2mm}5.2276$\times 10^{-7}$&\hspace{-2mm}5.1445$\times 10^{-7}$&\hspace{-2mm}5.0661$\times 10^{-7}$&\hspace{-2mm}4.9922$\times 10^{-7}$&\hspace{-2mm}4.9227$\times 10^{-7}$&\hspace{-2mm}4.8572$\times 10^{-7}$\\
28&\hspace{-2mm}3.6566$\times 10^{-7}$&\hspace{-2mm}3.6030$\times 10^{-7}$&\hspace{-2mm}3.5522$\times 10^{-7}$&\hspace{-2mm}3.5041$\times 10^{-7}$&\hspace{-2mm}3.4586$\times 10^{-7}$&\hspace{-2mm}3.4156$\times 10^{-7}$&\hspace{-2mm}3.3751$\times 10^{-7}$\\
30&\hspace{-2mm}2.5830$\times 10^{-7}$&\hspace{-2mm}2.5492$\times 10^{-7}$&\hspace{-2mm}2.5170$\times 10^{-7}$&\hspace{-2mm}2.4865$\times 10^{-7}$&\hspace{-2mm}2.4575$\times 10^{-7}$&\hspace{-2mm}2.4300$\times 10^{-7}$&\hspace{-2mm}2.4040$\times 10^{-7}$\\
40&\hspace{-2mm}6.0949$\times 10^{-8}$&\hspace{-2mm}6.0447$\times 10^{-8}$&\hspace{-2mm}5.9964$\times 10^{-8}$&\hspace{-2mm}5.9502$\times 10^{-8}$&\hspace{-2mm}5.9058$\times 10^{-8}$&\hspace{-2mm}5.8632$\times 10^{-8}$&\hspace{-2mm}5.8225$\times 10^{-8}$\\
50&\hspace{-2mm}1.9959$\times 10^{-8}$&\hspace{-2mm}1.9844$\times 10^{-8}$&\hspace{-2mm}1.9732$\times 10^{-8}$&\hspace{-2mm}1.9625$\times 10^{-8}$&\hspace{-2mm}1.9521$\times 10^{-8}$&\hspace{-2mm}1.9421$\times 10^{-8}$&\hspace{-2mm}1.9324$\times 10^{-8}$\\
60&\hspace{-2mm}8.0278$\times 10^{-9}$&\hspace{-2mm}7.9930$\times 10^{-9}$&\hspace{-2mm}7.9592$\times 10^{-9}$&\hspace{-2mm}7.9264$\times 10^{-9}$&\hspace{-2mm}7.8947$\times 10^{-9}$&\hspace{-2mm}7.8640$\times 10^{-9}$&\hspace{-2mm}7.8342$\times 10^{-9}$\\
70&\hspace{-2mm}3.7189$\times 10^{-9}$&\hspace{-2mm}3.7062$\times 10^{-9}$&\hspace{-2mm}3.6939$\times 10^{-9}$&\hspace{-2mm}3.6819$\times 10^{-9}$&\hspace{-2mm}3.6702$\times 10^{-9}$&\hspace{-2mm}3.6589$\times 10^{-9}$&\hspace{-2mm}3.6479$\times 10^{-9}$\\
80&\hspace{-2mm}1.9100$\times 10^{-9}$&\hspace{-2mm}1.9047$\times 10^{-9}$&\hspace{-2mm}1.8996$\times 10^{-9}$&\hspace{-2mm}1.8945$\times 10^{-9}$&\hspace{-2mm}1.8896$\times 10^{-9}$&\hspace{-2mm}1.8849$\times 10^{-9}$&\hspace{-2mm}1.8802$\times 10^{-9}$\\
90&\hspace{-2mm}1.0613$\times 10^{-9}$&\hspace{-2mm}1.0588$\times 10^{-9}$&\hspace{-2mm}1.0564$\times 10^{-9}$&\hspace{-2mm}1.0541$\times 10^{-9}$&\hspace{-2mm}1.0518$\times 10^{-9}$&\hspace{-2mm}1.0496$\times 10^{-9}$&\hspace{-2mm}1.0474$\times 10^{-9}$\\
100&\hspace{-2mm}6.2743$\times 10^{-10}$&\hspace{-2mm}6.2620$\times 10^{-10}$&\hspace{-2mm}6.2500$\times 10^{-10}$&\hspace{-2mm}6.2382$\times 10^{-10}$&\hspace{-2mm}6.2267$\times 10^{-10}$&\hspace{-2mm}6.2155$\times 10^{-10}$&\hspace{-2mm}6.2045$\times 10^{-10}$\\
110&\hspace{-2mm}3.9001$\times 10^{-10}$&\hspace{-2mm}3.8935$\times 10^{-10}$&\hspace{-2mm}3.8870$\times 10^{-10}$&\hspace{-2mm}3.8807$\times 10^{-10}$&\hspace{-2mm}3.8745$\times 10^{-10}$&\hspace{-2mm}3.8685$\times 10^{-10}$&\hspace{-2mm}3.8626$\times 10^{-10}$\\
120&\hspace{-2mm}2.5268$\times 10^{-10}$&\hspace{-2mm}2.5230$\times 10^{-10}$&\hspace{-2mm}2.5194$\times 10^{-10}$&\hspace{-2mm}2.5158$\times 10^{-10}$&\hspace{-2mm}2.5123$\times 10^{-10}$&\hspace{-2mm}2.5088$\times 10^{-10}$&\hspace{-2mm}2.5054$\times 10^{-10}$\\
130&\hspace{-2mm}1.6949$\times 10^{-10}$&\hspace{-2mm}1.6927$\times 10^{-10}$&\hspace{-2mm}1.6905$\times 10^{-10}$&\hspace{-2mm}1.6884$\times 10^{-10}$&\hspace{-2mm}1.6863$\times 10^{-10}$&\hspace{-2mm}1.6842$\times 10^{-10}$&\hspace{-2mm}1.6822$\times 10^{-10}$\\
140&\hspace{-2mm}1.1711$\times 10^{-10}$&\hspace{-2mm}1.1697$\times 10^{-10}$&\hspace{-2mm}1.1683$\times 10^{-10}$&\hspace{-2mm}1.1670$\times 10^{-10}$&\hspace{-2mm}1.1657$\times 10^{-10}$&\hspace{-2mm}1.1645$\times 10^{-10}$&\hspace{-2mm}1.1632$\times 10^{-10}$\\
150&\hspace{-2mm}8.3001$\times 10^{-11}$&\hspace{-2mm}8.2914$\times 10^{-11}$&\hspace{-2mm}8.2829$\times 10^{-11}$&\hspace{-2mm}8.2745$\times 10^{-11}$&\hspace{-2mm}8.2662$\times 10^{-11}$&\hspace{-2mm}8.2581$\times 10^{-11}$&\hspace{-2mm}8.2501$\times 10^{-11}$\\
\hline \hline
\end{tabular}
\label{tab:energy_flux}
\end{center}
\end{table}

\begin{table}[htbp]
\begin{center}
\caption{Relative error of the energy flux between Ref. \citen{TN} and our results for $r_0=10M,q=0$.}
\begin{tabular}{cc|ccc}
\hline \hline 
$\ell$&$\mid m\mid$&{\rm Tagoshi and Nakamura}&{\rm This paper}&{\rm Absolute values of relative error}\\ \hline 
$2$&$1$&$1.9316093511566875\times 10^{-7}$&$1.9316093511566907\times 10^{-7}$&$1.64\times 10^{-15}$\\
$2$&$2$&$5.3687954791021136\times 10^{-5}$&$5.3687954791021361\times 10^{-5}$&$4.20\times 10^{-15}$\\
$3$&$1$&$5.7148989126147667\times 10^{-10}$&$5.7148989126147834\times 10^{-10}$&$2.91\times 10^{-15}$\\
$3$&$2$&$4.7959164615902625\times 10^{-8}$&$4.7959164615902543\times 10^{-8}$&$1.71\times 10^{-15}$\\
$3$&$3$&$6.4260827562472322\times 10^{-6}$&$6.4260827562472412\times 10^{-6}$&$1.39\times 10^{-15}$\\
$4$&$1$&$1.4575856422971336\times 10^{-13}$&$1.4575856422971290\times 10^{-13}$&$3.19\times 10^{-15}$\\
$4$&$2$&$5.2622453089592304\times 10^{-10}$&$5.2622453089592954\times 10^{-10}$&$1.24\times 10^{-14}$\\
$4$&$3$&$8.7787575252149415\times 10^{-9}$&$8.7787575252150748\times 10^{-9}$&$1.52\times 10^{-14}$\\
$4$&$4$&$9.5396003948519482\times 10^{-7}$&$9.5396003948518797\times 10^{-7}$&$7.18\times 10^{-15}$\\
$5$&$1$&$2.3676371874495427\times 10^{-16}$&$2.3676371874495484\times 10^{-16}$&$2.40\times 10^{-15}$\\
$5$&$2$&$3.8193532371989606\times 10^{-13}$&$3.8193532371989253\times 10^{-13}$&$9.24\times 10^{-15}$\\
$5$&$3$&$1.8291013252282956\times 10^{-10}$&$1.8291013252283069\times 10^{-10}$&$6.20\times 10^{-15}$\\
$5$&$4$&$1.4921162748528251\times 10^{-9}$&$1.4921162748527967\times 10^{-9}$&$1.91\times 10^{-14}$\\
$5$&$5$&$1.5241547645798743\times 10^{-7}$&$1.5241547645799033\times 10^{-7}$&$1.90\times 10^{-14}$\\
$6$&$1$&$3.5977953599117842\times 10^{-20}$&$3.5977953599117314\times 10^{-20}$&$1.47\times 10^{-14}$\\
$6$&$2$&$1.9763689535200328\times 10^{-15}$&$1.9763689535200461\times 10^{-15}$&$6.73\times 10^{-15}$\\
$6$&$3$&$2.1238827476369010\times 10^{-13}$&$2.1238827476368560\times 10^{-13}$&$2.12\times 10^{-14}$\\
$6$&$4$&$4.6633398847411170\times 10^{-11}$&$4.6633398847412050\times 10^{-11}$&$1.89\times 10^{-14}$\\
$6$&$5$&$2.4746386947271652\times 10^{-10}$&$2.4746386947272387\times 10^{-10}$&$2.97\times 10^{-14}$\\
$6$&$6$&$2.5182131568101545\times 10^{-8}$&$2.5182131568101642\times 10^{-8}$&$3.82\times 10^{-15}$\\
$7$&$1$&$3.2913629491589289\times 10^{-23}$&$3.2913629491588661\times 10^{-23}$&$1.91\times 10^{-14}$\\
$7$&$2$&$9.0841508908488114\times 10^{-19}$&$9.0841508908487538\times 10^{-19}$&$6.34\times 10^{-15}$\\
$7$&$3$&$2.0373627509685863\times 10^{-15}$&$2.0373627509686016\times 10^{-15}$&$7.53\times 10^{-15}$\\
$7$&$4$&$6.9940936502071495\times 10^{-14}$&$6.9940936502074081\times 10^{-14}$&$3.70\times 10^{-14}$\\
$7$&$5$&$1.0340989127935083\times 10^{-11}$&$1.0340989127934911\times 10^{-11}$&$1.66\times 10^{-14}$\\
$7$&$6$&$4.0679948091711724\times 10^{-11}$&$4.0679948091710895\times 10^{-11}$&$2.04\times 10^{-14}$\\
$7$&$7$&$4.2345226712846746\times 10^{-9}$&$4.2345226712847762\times 10^{-9}$&$2.40\times 10^{-14}$\\
\hline \hline
\end{tabular}
\label{tab:flux-compare_TN}
\end{center}
\end{table}

\begin{table}[htbp]
\begin{center}
\caption{Relative error of the energy flux between Kennefick's results\cite{GK2} 
and our results for $r_0=3M,q=0.998$.}
\begin{tabular}{cc|ccc}
\hline \hline 
$\ell$&$\mid m\mid$&{\rm Kennefick}&{\rm This paper}&{\rm Absolute values of relative error}\\ \hline 
$2$&$1$&$6.328361227\times 10^{-6}$&$6.3283737145864869\times 10^{-6}$&$1.97\times 10^{-6}$\\
$2$&$2$&$7.335124682\times 10^{-3}$&$7.3350910900390372\times 10^{-3}$&$4.58\times 10^{-6}$\\
$3$&$1$&$3.451784675\times 10^{-8}$&$3.4517924227877552\times 10^{-8}$&$2.24\times 10^{-6}$\\
$3$&$2$&$5.645895586\times 10^{-6}$&$5.6458902861747337\times 10^{-6}$&$9.39\times 10^{-7}$\\
$3$&$3$&$2.208481088\times 10^{-3}$&$2.2084763695700243\times 10^{-3}$&$2.14\times 10^{-6}$\\
$4$&$1$&$8.842632418\times 10^{-12}$&$8.8426341846420484\times 10^{-12}$&$2.00\times 10^{-7}$\\
$4$&$2$&$9.650747825\times 10^{-8}$&$9.6507761069096606\times 10^{-8}$&$2.93\times 10^{-6}$\\
$4$&$3$&$3.157205166\times 10^{-6}$&$3.1572084515052529\times 10^{-6}$&$1.04\times 10^{-6}$\\
$4$&$4$&$7.897993006\times 10^{-4}$&$7.8979614897263311\times 10^{-4}$&$3.99\times 10^{-6}$\\
$5$&$1$&$2.226601380\times 10^{-14}$&$2.2266003587317006\times 10^{-14}$&$4.59\times 10^{-7}$\\
$5$&$2$&$7.623615371\times 10^{-11}$&$7.6236253541558626\times 10^{-11}$&$1.31\times 10^{-6}$\\
$5$&$3$&$9.637713992\times 10^{-8}$&$9.6377094209411545\times 10^{-8}$&$4.74\times 10^{-7}$\\
$5$&$4$&$1.524036288\times 10^{-6}$&$1.5240423964556889\times 10^{-6}$&$4.01\times 10^{-6}$\\
$5$&$5$&$3.044015502\times 10^{-4}$&$3.0439980326075501\times 10^{-4}$&$5.74\times 10^{-6}$\\
$6$&$1$&$3.850986300\times 10^{-18}$&$3.8509863211419323\times 10^{-18}$&$5.49\times 10^{-9}$\\
$6$&$2$&$5.736053080\times 10^{-13}$&$5.7360410503468190\times 10^{-13}$&$2.10\times 10^{-6}$\\
$6$&$3$&$1.315574802\times 10^{-10}$&$1.3155762424734953\times 10^{-10}$&$1.09\times 10^{-6}$\\
$6$&$4$&$6.788432814\times 10^{-8}$&$6.7884404091806854\times 10^{-8}$&$1.12\times 10^{-6}$\\
$6$&$5$&$6.900330654\times 10^{-7}$&$6.9003809076206216\times 10^{-7}$&$7.28\times 10^{-6}$\\
$6$&$6$&$1.219262418\times 10^{-4}$&$1.2192534271212692\times 10^{-4}$&$7.37\times 10^{-6}$\\
$7$&$1$&$5.063200000\times 10^{-21}$&$5.0631852897823082\times 10^{-21}$&$2.91\times 10^{-6}$\\
$7$&$2$&$3.106010424\times 10^{-16}$&$3.1060171837611518\times 10^{-16}$&$2.18\times 10^{-6}$\\
$7$&$3$&$1.769856886\times 10^{-12}$&$1.7698525192439268\times 10^{-12}$&$2.47\times 10^{-6}$\\
$7$&$4$&$1.284621326\times 10^{-10}$&$1.2846205001945888\times 10^{-10}$&$6.43\times 10^{-7}$\\
$7$&$5$&$4.051197224\times 10^{-8}$&$4.0511763345368075\times 10^{-8}$&$5.16\times 10^{-6}$\\
$7$&$6$&$3.023851152\times 10^{-7}$&$3.0238471412501713\times 10^{-7}$&$1.33\times 10^{-6}$\\
$7$&$7$&$4.993950576\times 10^{-5}$&$4.9939392339609413\times 10^{-5}$&$2.27\times 10^{-6}$\\
\hline \hline
\end{tabular}
\label{tab:flux-compare_kenn}
\end{center}
\end{table}

\section{Summary and discussion}
\label{sec:summary}
In this paper, we described numerical methods to compute the homogeneous
solutions of the Teukolsky equation. 
We used the MST formalism, in which the homogeneous solutions of the 
Teukolsky equation are expressed in terms of series of 
hypergeometric functions and Coulomb wave functions. 
We found that the renormalized angular momentum $\nu$ can be found 
only for a limited values of $M\omega$. 
When $M\omega$ becomes large, $\nu$ approaches $\ell-1/2$, and 
over a certain range of values of $M\omega$, which depends on $\ell,m$ and $q$, 
we could not find real $\nu$. 
Our preliminary investigations suggest that $\nu$ becomes complex 
when $M\omega$ becomes sufficiently large. However, because further investigation
is needed to confirm this, we continue to work on this point. 

In the region of $M\omega$ in which we can find real $\nu$, 
we found that the convergence of the series of 
hypergeometric functions or Coulomb wave functions is very rapid. 
This is because the expansion coefficients $\{f_n^\nu\}$,
which constitute a minimal solution of the three-term recurrence relation,
decrease very rapidly as $|n|$ becomes large. 
Thus, we concluded that the MST formalism is a powerful method 
to compute the homogeneous solutions numerically. 
By comparing the numerical data for the gravitational wave luminosity
emitted to infinity in the case that a particle display circular orbits
in the equatorial plane with the results of previous works,
we estimated the accuracy of our code to be about 13--14 significant
figures in the double precision computation. 
This accuracy will be sufficient as a Green function of the Teukolsky equation
in the computation of the templates for the data analysis of LISA. 
Currently, the accuracy is limited not by the MST formalism itself 
but by the accuracy 
of the evaluation of the hypergeometric functions.

In the near future, we will investigate the question of whether
complex $\nu$ exist in order to broaden the 
region of $M\omega$, in which this method can be applied. 
We also hope to improve the accuracy of the evaluation of the 
hypergeometric functions and Coulomb wave functions 
and to apply this method to the computation of 
gravitational waves from a compact star in the case of more generic orbits around 
a supermassive black hole. 

\section*{Acknowledgements}
We thank D. Kennefick for kindly providing us his numerical data 
of the gravitational wave luminosity computed with the code 
developed in his work \cite{GK}. 
We thank M. Sasaki for useful discussions and continuous encouragement. 
We also thank F. Takahara for continuous encouragement. 
H. T. thanks K. S. Thorne for continuous encouragement and his hospitality
while staying at California Institute of Technology, where a part of
this work was done, as a Zaigai Kenkyuin of Monbukagaku-sho. 
This work was supported in part by Monbukagaku-sho Grant-in-Aid
for Scientific Research (Nos. 14047214, 12640269 and 16540251). 

\appendix

\section{Teukolsky Formalism}
\label{sec:Teukolsky}
\subsection{Teukolsky equation}

In terms of the Boyer-Lindquist coordinates $(t,r,\theta,\phi)$, 
the metric of a Kerr black hole is expressed as 
\begin{eqnarray}
ds^2=&-&\frac{\Delta}{\Sigma}(dt-a\sin ^2 \theta d \phi)^2 +\frac{\sin ^2 
\theta}{\Sigma}\left[(r^2+a^2)d\phi-a\,dt\right]^2 \nonumber \\
&+&\frac{\Sigma}{\Delta}dr^2+\Sigma d\theta^2\equiv g_{\a \b}dx^{\a}dx^{\b}.
\end{eqnarray}
where $\Sigma=r^2+a^2\cos^2\theta$ and $\Delta=r^2-2Mr+a^2$. 
In the Teukolsky formalism \cite{Teukolsky}, the gravitational perturbations 
of a Kerr black hole are described by the Newman-Penrose quantity 
$\psi_4=-C_{\a\b\c\d}n^{\a}\ol{m}^{\b}n^{\c}\ol{m}^{\d}$, where
$C_{\a\b\c\d}$ is the Weyl tensor and 
\begin{eqnarray}
n ^{\alpha }&=&\left[(r^2+a^2),-\Delta ,0,a\right]/(2\Sigma)\nonumber ,\\
m ^{\alpha }&=&\left[i a\sin \theta,0,1,i/\sin \theta \right]/(\sqrt{2}(r+i a 
\sin \theta))\nonumber ,\\
\ol{m} ^{\alpha }&=&\left[-i a\sin \theta,0,1,-i/\sin \theta \right]/(\sqrt{2}(r-i a 
\sin \theta)).
\end{eqnarray}

The Weyl curvature component $\psi_{4}$ contains all the information 
regarding the 
gravitational radiation. Teukolsky showed that  
if we carry out Fourier-harmonic decomposition of $\rho^{-4}\psi_{4}$, 
we can separate the 
Teukolsky equation into a radial part and an angular part, or
\begin{eqnarray}
\rho^{-4}\psi_4(t,r,\theta,\phi)&=&\sum_{\ell, m}\int d\omega 
e^{-i\omega t+i m \varphi} \ _{-2}S_{\ell m}^{a\omega}(\theta)
R_{\ell m\omega}(r),\nonumber \\
4\pi\Sigma\hat{T}&=&\sum_{\ell, m}\int d\omega 
e^{-i\omega t+i m \varphi} \ _{-2}S_{\ell m}^{a\omega}(\theta)
T_{\ell m\omega}(r),
\end{eqnarray}
where $\rho= (r -ia\cos\theta)^{-1}$.

The radial function $R_{lm\omega}(r)$ and the angular function 
$_{-2}S_{lm}^{a\omega}(\theta)$ satisfy the following Teukolsky equations: 
\begin{eqnarray}
\label{eq:radial-Teukolsky}
\Delta^2{d\over dr}\left({1\over \Delta}{dR_{\ell m\omega}\over dr}
\right)
-V(r) R_{\ell m\omega}=T_{\ell m\omega} ,
\end{eqnarray}
\begin{eqnarray}
\label{eq:ang-Teukolsky}
\left [\frac{1}{\sin \theta }\frac{d}{d\theta}\left\{\sin \theta\frac{d}
{d\theta}\right\}-a^2 \omega ^2 \sin ^2\theta-\frac{(m-2\cos \theta)^2}
{\sin ^2\theta}\right.\nonumber \\
+4 a \omega \cos \theta -2 +2ma\omega +\lambda \biggr]\ _{-2}S_{\ell m}^{a\omega}=0.
\end{eqnarray}
The potential is given by 
\begin{eqnarray}
V(r) = -{K^2 + 4i(r-M)K\over\Delta} + 8i\omega r + \lambda,
\end{eqnarray}
where $K=(r^2+a^2)\omega-ma$ and $\lambda$ is the 
eigenvalue of $_{-2}S^{a\omega}_{\ell m}(\theta)$. 
The angular function $_{-2}S^{a\omega}_{\ell m}(\theta)$ is called a 
spin-weighted spheroidal harmonic with spin weight $-2$. 
It is usually normalized as
\begin{eqnarray}
\label{eq:normalSp}
\int _{0}^{\pi}\left |\ _{-2}S_{\ell m}^{a\omega}\right |^2\sin \theta d\theta=1.
\end{eqnarray}

In the case of a Kerr black hole, analytic form of 
the spin-weighted spheroidal harmonics and the eigenvalue $\lambda$ are not known.
However, in the case of a Schwarzschild black hole, 
the spin-weighted spheroidal harmonics reduce to 
the spin-weighted spherical harmonics, and its eigenvalue is 
$\lambda=\ell(\ell+1)-s(s+1)$ \cite{Press}.

We solve the radial Teukolsky equation using the Green function method. 
For this purpose, we introduce two kinds of 
the homogeneous solutions of the radial Teukolsky equation,
\begin{eqnarray}
\label{eq:Rin-asymp}
R_{lm\omega}^{{\rm in}}&\rightarrow&\left\{
\begin{array}{cc}
B_{lm\omega}^{{\rm trans}}\Delta^2 e^{-ikr*}&{\rm for}\ r\rightarrow r_{+},\\
r^3 B_{lm\omega}^{{\rm ref}}e^{i\omega r*}+r^{-1}B_{lm\omega}^{{\rm inc}}e^{-i\z r*}&{\rm for}\ r\rightarrow +\infty,
\end{array}
\right.\\
\label{eq:Rup-asymp}
R_{lm\omega}^{{\rm up}}&\rightarrow&\left\{
\begin{array}{cc}
C_{lm\omega}^{{\rm up}}e^{ikr*}+\Delta^2 C_{lm\omega}^{{\rm ref}}e^{-ikr*}&{\rm for}\ r\rightarrow r_{+},\\
r^3 C_{lm\omega}^{{\rm trans}}e^{i\omega r*}&{\rm for}\ r\rightarrow +\infty,
\end{array}
\right.
\end{eqnarray}
where $k=\omega-ma/2Mr_{+}$ and $r^{*}$ is the tortoise coordinate defined by
\begin{eqnarray}
r^*&=&\int \frac{{\rm d}r^*}{{\rm d}r}{\rm d}r,\nonumber \\
&=&\int \frac{r^2+a^2}{\Delta}{\rm d}r,\nonumber \\
&=&r+\frac{2Mr_{+}}{r_{+}-r_{-}}\ln\frac{r-r_{+}}{2M}-\frac{2Mr_{-}}{r_{+}-r_{-}}\ln\frac{r-r_{-}}{2M}.
\end{eqnarray}
Here $r_{\pm}=M\pm\sqrt{M^2-a^2}$. 

Then, a solution of the radial 
Teukolsky equation that is ingoing at the horizon and 
outgoing at the infinity can be written as
\begin{eqnarray*}
R_{\ell m \omega}(r)&=&\frac{1}{W_{lm\omega}}\left\{R^{\rm up}_{\ell m\omega}(r)
\int^{r}_{r_{+}} dr' \frac{T_{\ell m\omega}(r')R^{\rm in}_{\ell m\omega}(r')}
{\Delta^{2}(r')} 
\right.
\nonumber\\
&&
\left.
+ R^{\rm in}_{\ell m\omega}(r)\int^{\infty}_{r} dr' 
\frac{T_{\ell m\omega}(r')R^{\rm up}_{\ell m\omega}(r')}
{\Delta^{2}(r')}\right\},
\label{Rfield}
\end{eqnarray*}
where $W_{lm\omega}$ is the Wronskian given by
\begin{equation}
W_{\ell m\omega}=
W[ \Delta^{-1/2} R^{\rm in}_{\ell m\omega}, \Delta^{-1/2} 
R^{\rm up}_{\ell m\omega} ]
=2 i \omega C^{\rm trans}_{\ell m\omega}
     B^{\rm inc}_{\ell m\omega}. 
\end{equation}

The asymptotic behavior at the horizon is given by 
\begin{eqnarray}
\label{eq:Rhorizon}
R_{\ell m\omega}(r\to r_+)
&\to&
{{B^{\rm trans}_{\ell m\omega} \Delta^2 e^{-i k r^*}} \over 
2i\omega C^{\rm trans}_{\ell m\omega}B^{\rm inc}_{\ell m\omega}}
\int^{\infty}_{r_+}dr' \frac{T_{\ell m\omega}(r')
	R^{\rm up}_{\ell m\omega}(r')}{\Delta^{2}(r')}
\nonumber\\
&&
\equiv \tilde{Z}^{\rm H}_{\ell m\omega} \Delta^2 e^{-i k r^*}. 
\end{eqnarray}
The asymptotic behavior at infinity is given by 
\begin{eqnarray}
\label{eq:Rinfty}
R_{\ell m\omega}(r\to\infty)
&\to&
{r^3e^{i\omega r^*} \over 2i\omega 
   B^{\rm inc}_{\ell m\omega}}
\int^{\infty}_{r_+}dr'{T_{\ell m\omega}(r') 
R^{\rm in}_{\ell m\omega}(r') 
\over\Delta^{2}(r')}
\equiv \tilde Z_{\ell m\omega}^{\infty}r^3e^{i\omega r^*}.
\end{eqnarray}

In this paper, we focus on the gravitational wave flux from a point particle 
in circular, equatorial orbits around a Kerr black hole. In this case, 
$\tilde Z_{lm\omega}$ in Eq. (\ref{eq:Rinfty}) takes the form
\begin{eqnarray}
\label{eq:discreteZ}
\tilde Z_{lm\omega}^{\infty}=
\sum_{n} \delta(\omega-\omega_{n})Z_{lm\omega_n}^{\infty},
\end{eqnarray}
where $\omega_n=m M^{1/2}/(r^{3/2}+a M^{1/2})$.

Then, the time-averaged flux (luminosity) radiated to infinity is given by
\begin{eqnarray}
\label{eq:fluxatinfty}
\frac{{\rm d}E}{{\rm d}t}=\sum _{l,m,n} 
\frac{\mid Z_{lm\omega_n}^{\infty}\mid^2}
{4\pi\omega_{n}^2}.
\end{eqnarray}

\subsection{Source term}

The source term of the Teukolsky equation, Eq. (\ref{eq:radial-Teukolsky}), 
is given by 
\begin{equation} 
T_{\ell m\omega}
 =4\int d\Omega dt\rho^{-5}{\bar \rho}^{-1}(B_2'+B_2'^*)
e^{-im\varphi+i\omega t}{_{-2}S^{a\omega}_{\ell m} \over
\sqrt{2\pi}},
\label{source}
\end{equation}
where $B_2'$ and $B_2^*$ are given by 
\begin{eqnarray}
B_2'&=&-{1 \over 2}\rho^8{\bar \rho}L_{-1}[\rho^{-4}L_0
(\rho^{-2}{\bar \rho}^{-1}T_{nn})]  \nonumber\\
&-&{1 \over 2\sqrt{2}}\rho^8{\bar \rho}\Delta^2 L_{-1}[\rho^{-4}
{\bar \rho}^2 J_+(\rho^{-2}{\bar \rho}^{-2}\Delta^{-1}
T_{{\bar m}n})], \nonumber \\
B_2'^*&=&-{1 \over 4}\rho^8{\bar \rho}\Delta^2 J_+[\rho^{-4}J_+
(\rho^{-2}{\bar \rho}T_{{\bar m}{\bar m}})] \nonumber\\
&-&{1 \over 2\sqrt{2}}\rho^8{\bar \rho}\Delta^2 J_+[\rho^{-4}
{\bar \rho}^2 \Delta^{-1} L_{-1}(\rho^{-2}{\bar \rho}^{-2}
T_{{\bar m}n})] \;, 
\label{eq:B2}
\end{eqnarray}
\begin{eqnarray}
{\cal L}_s&=&\partial_{\theta}+{m \over \sin\theta} -a\omega\sin\theta +s\cot\theta, 
\\
\nonumber \\
J_+&=&\partial_r+{iK/\Delta}\ ;\ \ K = (r^2+a^2)\omega - m a,\nonumber
\end{eqnarray}
and $T_{nn}$, $T_{\ol{m}n}$, $T_{\ol{m}\ol{m}}$ are the tetrad components of 
the energy momentum tensor ($T_{nn}=T_{\m\n}n^{\m}n^{\n}$ etc.). Here the bar denotes 
the complex conjugation.

We consider $T_{\mu\nu}$ of a monopole particle of mass $\mu$. In this
case, the energy 
momentum tensor takes the form
\begin{eqnarray*}
T^{\mu\nu}
&=& \mu \int d\tau\frac{d\,z^{\mu}}{d\tau}\frac{d\,z^{\nu}}{d\tau} 
\frac{\delta^{(4)}(x-z(\tau))}{\sqrt{-g}}\nonumber ,\\
&=& \mu \frac{u^{\mu}u^{\nu}}{\Sigma \sin\theta u^{t} } \delta(r -r(t))
\delta(\theta -\theta(t)) \delta(\phi -\phi(t)) \;,
\label{emtensor}
\end{eqnarray*}
where $ u^{\mu}= dz^{\mu}/d\tau $ and $z^{\mu}=(t,r(t),\theta(t),\phi(t))$
is a geodesic trajectory and $\tau=\tau(t)$ is the proper time along this 
geodesic. The geodesic equations in the Kerr geometry are given by 
\begin{eqnarray}
\Sigma\frac{{\rm d} t}{{\rm d} \tau}&=&\frac{r^{2}+a^{2}}{\Delta}
\left[E(r^{2}+a^{2}) - a L_{z}\right] - a\left[a E\sin^{2}\theta-L_{z}\right],
\nonumber \\
\Sigma\frac{{\rm d} r}{{\rm d} \tau}&=&\pm\sqrt{R},\nonumber \\
\Sigma\frac{{\rm d} \theta}{{\rm d} \tau}&=&\pm\sqrt{\Theta},\nonumber \\
\Sigma\frac{{\rm d} \phi}{{\rm d} \tau}&=&\frac{a}{\Delta}\left[E(r^{2}
+a^{2}) - a L_{z}\right]- a E + \frac{L_{z}}{\sin^{2}\theta},
\label{eq:geodesic}
\end{eqnarray}
where $E$ and $L_{z}$ are the energy and the z-component of the angular 
momentum of a test particle, respectively. Also, we have 
\begin{eqnarray}
  R      &=& \left[(r^2+a^2)E-a L_{z}\right]^2
             - \Delta[r^2+(L_{z}-a E)^2+C]\nonumber , \\
  \Theta &=& C - \left[(1-E^2)a^2
             + \frac{L_{z}^2}{\sin^2\theta}\right]\cos^2\theta
	     \nonumber ,
\end{eqnarray}
where $C$ is the Carter constant of a test particle.
Note that $E, L_z$ and $C$ are measured in units of $\mu$.
If they are expressed in the standard units, 
$E, L_z$ and $C$ in the above equations must be replaced by 
$E/\mu$, $L_z/\mu$ and $C/\mu^2$. 

Using Eq. (\ref{eq:geodesic}), we obtain the tetrad components of the energy 
momentum tensor as
\begin{eqnarray}
T_{nn}&=&\mu{C_{nn} \over \sin\theta}
\delta(r-r(t)) \delta(\theta-\theta(t)) \delta(\phi-\phi(t)),
\\
\nonumber \\
\nonumber \\
T_{{\bar m}n}&=&\mu{C_{{\bar m} n} \over \sin\theta}
\delta(r-r(t)) \delta(\theta-\theta(t)) \delta(\phi-\phi(t)),
\\
\nonumber \\
\nonumber \\
T_{{\bar m}{\bar m}}&=&\mu{C_{{\bar m}{\bar m}} \over \sin\theta}
\delta(r-r(t)) \delta(\theta-\theta(t)) \delta(\phi-\phi(t)) \;,
\end{eqnarray}
where
\begin{eqnarray}
C_{nn}&=&{1\over 4\Sigma^3 u^{\rm t}}\left[E(r^2+a^2)-a L_{z}
+\Sigma \frac{{\rm d}r}{{\rm d}\tau} \right]^2,
\nonumber\\
C_{{\bar m}n}&=& -{\rho \over 2\sqrt{2}\Sigma^2 u^{\rm t}}
\left[E(r^2+a^2)-aL_{z} +\Sigma \frac{{\rm d}r}{{\rm d}\tau} \right]
\left[i\sin\theta\Bigl(aE-{L_{z} \over \sin^2\theta}\Bigr) +
\Sigma\frac{{\rm d}\theta}{{\rm d}\tau} \right], 
\nonumber \\
C_{{\bar m}{\bar m}}&=& {\rho^2 \over 2\Sigma u^{\rm t}}
\left[i\sin\theta \Bigl(aE-{L_z \over \sin^2\theta}\Bigr) + 
\Sigma\frac{{\rm d}\theta}{{\rm d}\tau} \right]^2 \;.
\label{Cq}
\end{eqnarray}
Substituting Eq. (\ref{eq:B2}) into Eq. (\ref{source}) and performing 
integration by parts, we obtain
\begin{eqnarray}
T_{\ell m\omega}&=&{4\mu\over\sqrt{2\pi}}\int^{\infty}_{-\infty}
dt\int_{0}^{\pi} d\theta e^{i\omega t-im\varphi(t)} \nonumber\\
&\times&
\Bigl[-{1 \over 2}L_1^{\dag} \bigl\{ \rho^{-4}L_2^{\dag}(\rho^3 S_{\ell 
m}^{a\omega})
\bigr\}
C_{nn}\rho^{-2}{\bar \rho}^{-1}\delta(r-r(t))
\delta(\theta-\theta(t)) \nonumber\\
&+&{\Delta^2 {\bar \rho}^2 \over \sqrt{2} \rho}
\bigl(L_2^{\dag} S_{\ell m}^{a\omega} + ia({\bar \rho}-\rho)\sin\theta 
S_{\ell m}^{a\omega}\bigr) 
\nonumber\\
&&\quad 
\times J_+ \bigl\{ C_{{\bar m}n}\rho^{-2}{\bar \rho}^{-
2}\Delta^{-1} \delta(r-r(t))\delta(\theta-\theta(t)) \bigr\} \nonumber\\
&+&{1 \over 2\sqrt{2} }
L_2^{\dag}\bigl\{ \rho^3 S_{\ell m}^{a\omega} ({\bar \rho}^2 \rho^{-
4})_{,r} 
\bigr\} C_{{\bar m}n}\Delta \rho^{-2}{\bar \rho}^{-2}
\delta(r-r(t))\delta(\theta-\theta(t)) \nonumber\\
&-&{1 \over 4}\rho^3 \Delta^2 S_{\ell m}^{a\omega} J_+\bigl\{\rho^{-4}
J_+\bigl({\bar \rho} \rho^{-2}C_{{\bar m}{\bar m}}
\delta(r-r(t))\delta(\theta-\theta(t))\bigr) \bigr\}
\Bigr],
\label{source2}\\
\nonumber
\end{eqnarray}
where 
\begin{equation}
{\cal L}_s^{\dag}=\partial_{\theta}-{m \over \sin\theta}
+a\omega\sin\theta+s\cot\theta \;,
\end{equation}
and $S_{\ell m}^{a\omega}$  denotes $_{-2}S_{\ell m}^{a\omega}(\theta)$ 
for simplicity.

The integration over $\theta$ can be performed directly. It yields
\begin{eqnarray}
T_{\ell m\omega}&=&\mu\int^{\infty}_{-\infty}dt 
e^{i\omega t-i m \varphi(t)}
\Delta^2\Bigl[(A_{nn0}+A_{{\bar m}n0}+
A_{{\bar m}{\bar m}0})\delta(r-r(t)) \nonumber\\
&+&\left\{(A_{{\bar m}n1}+A_{{\bar m}{\bar m}1})
\delta(r-r(t))\right\}_{,r}
+\left\{A_{{\bar m}{\bar m}2}
\delta(r-r(t))\right\}_{,rr}\Bigr]_{\theta= \theta(t)} \;,
\label{eq:source3} 
\end{eqnarray}
where
\begin{eqnarray}
A_{nn0}&=&{-2 \over \sqrt{2\pi}\Delta^2}
C_{nn}\rho^{-2}{\bar \rho}^{-1}
L_1^+\{\rho^{-4}L_2^+(\rho^3 S_{\ell m}^{a\omega})\},
\\
\nonumber \\
\nonumber \\
A_{{\bar m}n0}&=&{2 \over \sqrt{\pi}\Delta} 
C_{{\bar m}n}\rho^{-3}
\Bigl[\left(L_2^+S_{\ell m}^{a\omega}\right)
\Bigl({iK \over \Delta}+\rho+{\bar \rho}\Bigr) 
\nonumber \\
& &\qquad-a\sin\theta(t) S_{\ell m}^{a\omega} {K \over \Delta}
({\bar \rho}-\rho)\Bigr], 
\\
A_{{\bar m}{\bar m}0}
&=&-{1 \over \sqrt{2\pi}}\rho^{-3}{\bar \rho}
C_{{\bar m}{\bar m}}S_{\ell m}^{a\omega}\Bigl[
-i\Bigl({K \over \Delta}\Bigr)_{,r}-{K^2 \over \Delta^2}+
2i\rho {K \over \Delta}\Bigr],
\\
A_{{\bar m}\,n\,1}&=&{2\over \sqrt{\pi}\Delta }
\rho^{-3}C_{{\bar m}n}
[L_2^{+}S_{\ell m}^{a\omega} +ia\sin\theta(t)({\bar \rho}-\rho)
S_{\ell m}^{a\omega}], 
\\
A_{{\bar m}{\bar m}1} &=&-{2 \over \sqrt{2\pi}} \rho^{-3}{\bar \rho}
C_{{\bar m}{\bar m}}S_{\ell m}^{a\omega}
\Bigl(i{K \over \Delta}+\rho\Bigr), 
\\
A_{{\bar m}{\bar m}2}
&=&-{1\over \sqrt{2\pi}}\rho^{-3}{\bar \rho}
C_{{\bar m}{\bar m}}S_{\ell m}^{a\omega}. 
\label{Aq} 
\end{eqnarray}
Inserting Eq. (\ref{eq:source3}) into Eqs. (\ref{eq:Rhorizon}) and 
(\ref{eq:Rinfty}), we obtain $Z_{\ell m\omega}^{\infty,H}$ as
\begin{eqnarray}
Z^{H}_{\ell m\omega} &=&
{\mu B^{\rm trans}_{lm\omega}\over2i\omega C^{\rm trans}_{lm\omega}
B^{\rm inc}_{lm\omega}}
\int^{\infty}_{-\infty}dt e^{i\omega t-i m \varphi(t)}
{\cal I}^{H}_{\ell m \omega}(r(t),\theta(t)) \;,
\label{Zh} 
\\ 
\nonumber \\
\nonumber \\
 Z^{\infty}_{\ell m\omega} &=&
{\mu\over2i\omega B^{\rm inc}_{lm\omega}}
\int^{\infty}_{-\infty}dt e^{i\omega t-i m \varphi(t)}
{\cal I}^{\infty}_{\ell m \omega}(r(t),\theta(t))  \;,
\label{Zinf}
\end{eqnarray}
where
\begin{eqnarray}
{\cal I}^{\rm H}_{\ell m\omega}&=&
\Bigl[R^{\rm up}_{\ell m\omega}\{A_{nn0}+A_{{\bar m}n0}
+A_{{\bar m}\,{\bar m}\,0}\} 
\nonumber\\
& &-{dR^{\rm up}_{\ell m\omega} \over dr}\{ A_{{\bar m}n1}
+A_{{\bar m}{\bar m}1}\}
 +{d^2 R^{\rm up}_{\ell m\omega} \over dr^2}
 A_{{\bar m}{\bar m}2}\Bigr]_{r=r(t),\theta=\theta(t)} \;,
\\
\nonumber \\
\nonumber \\
{\cal I}^{\rm \infty}_{\ell m\omega}&=&
\Bigl[R^{\rm in}_{\ell m\omega}\{A_{nn0}+A_{{\bar m}n0}
+A_{{\bar m}{\bar m}0}\}
\nonumber\\
& &-{dR^{\rm in}_{\ell m\omega} \over dr}\{ A_{{\bar m}n1}
+A_{{\bar m}{\bar m}1}\}
 +{d^2 R^{\rm in}_{\ell m\omega} \over dr^2}
 A_{{\bar m}{\bar m}2}\Bigr]_{r=r(t),\theta=\theta(t)} \;.
\label{Iotas}
\end{eqnarray}

\section{The Spin-Weighted Spheroidal Harmonics}
\label{sec:SWSH}

In this section, we review the formalism to represent the spin-weighted spheroidal
harmonics in a series of Jacobi polynomials based on Ref.~\citen{Fackerell}. 
We also discuss a method to implement this formalism in the numerical 
computation. 

We first transform the angular Teukolsky equation as
\begin{eqnarray}
&&\left[(1-x^{2})\frac{d^2}{dx^{2}}-2x\frac{d}{dx}+{\xi}^{2} x^{2}
-\frac{m^{2}+s^{2}+2msx}{1-x^{2}}-2s\xi x+E \right]\ _{s}S_{lm}^{a\omega}(x)=0,
\nonumber\\
&&\label{eq:Sphe diff}
\end{eqnarray}
where $\xi = a\,\omega, x = \cos\theta$
and $E=\lambda+s(s+1)-a^{2} \omega^{2}+2\,a\,m\, \omega$.

The angular function $_{s}S_{lm}^{a\omega}(x)$ is called the
spin-weighted spheroidal harmonics. Equation (\ref{eq:Sphe diff}) 
is a Sturm-Liouville type eigenvalue equation with the boundary conditions 
of the regularity at $x=\pm 1$. The eigenvalue $E$ depends on $\ell$ 
for fixed parameters $s$, $m$ and $a\omega$. Therefore we express the 
eigenvalue as $_{s}E_{l}^{m}(\xi)$.
When $a\omega=0$, $_{s}S_{lm}^{a\omega}(x)$ is reduced to the spin-weighted spherical 
harmonics, and the eigenvalue $_{s}E_{l}^{m}(\xi)$ becomes $l(l+1)$ \cite{Press}.

The differential equation Eq.~(\ref{eq:Sphe diff}) has singularities at
$x=\pm 1$ and $\infty$. We transform the angular function as 
\begin{eqnarray}
_{s}S_{lm}^{a\omega}(x)=e^{\xi x}\left(\frac{1-x}{2}\right)^{\frac{\alpha}{2}}
\left(\frac{1+x}{2}\right)^{\frac{\beta}{2}}\, _{s}U_{lm}(x),
\label{eq:SpheU}
\end{eqnarray}
where $\alpha = |m+s|$ and $\beta = |m-s|$. Then, Eq. (\ref{eq:Sphe diff}) 
becomes
\begin{eqnarray}
\label{eq:proto-Jacobi}
&&(1-x^{2})\,_{s}U_{lm}''(x)+\left[\beta-\alpha-(2+\alpha+\beta)x\right]\,_{s}U_{lm}'(x)
\nonumber\\
&&\quad
+\left[\,_{s}E_{l}^{m}(\xi)-\frac{\alpha+\beta}{2}\left(\frac{\alpha+\beta}{2}+
1\right)\right]\,_{s}U_{lm}(x)\nonumber \\
&&\quad
=\xi\left[-2(1-x^{2})\,_{s}U_{lm}'(x)+(\alpha+\beta+2s+2)x\,_{s}U_{lm}(x)
\right.
\nonumber\\
&&\quad\quad\left. 
-(\xi+\beta-\alpha)\,_{s}U_{lm}(x)\right].
\end{eqnarray}
When $\xi=0$, the right-hand side of Eq. (\ref{eq:proto-Jacobi}) is zero, 
while the left-hand side becomes the differential equation satisfied
by the Jacobi polynomials, 
\begin{eqnarray}
&&(1-x^{2})\,P_{r}^{(\alpha,\beta)}{}^{''}(x)
+\left[\beta-\alpha-(\alpha+\beta+2)x\right]\,P_{r}^{(\alpha,\beta)}{}^{'}(x)
\nonumber\\
&&
\quad +r(r+\alpha+\beta+1)\,P_{r}^{(\alpha,\beta)}(x)=0.
\label{eq:Jacobi}
\end{eqnarray}
In this limit, the eigenvalue $_{s}E_{\ell}^{m}$ in the equation 
(\ref{eq:proto-Jacobi}) becomes $\ell(\ell+1)$, 
where $\ell=r+(\alpha+\beta)/2=r+{\rm max}(\mid m\mid ,\mid s\mid )$.

In this paper, we use the Jacobi polynomials defined using the Rodrigue's formula by
\begin{eqnarray}
P_{n}^{(\alpha,\beta)}(x)=\frac{(-1)^{n}}{2^{n}\,n!}(1-x)^{-\alpha}(1+x)^
{-\beta}\left(\frac{d}{dx}\right)^{n}\left[(1-x)^{\alpha+n}(1+x)^{\beta+n}
\right].
\end{eqnarray}

Now, we expand $_{s}U_{lm}(x)$ in a series of Jacobi polynomials: 
\begin{eqnarray}
\label{eq:Jacobi-series}
_{s}U_{lm}(x)=\sum_{n=0}^{\infty}\,_{s}A_{lm}^{(n)}(\xi)\,P_{n}^{(\alpha,\beta)}(x).
\end{eqnarray}
The expansion coefficients $_{s}A_{lm}^{(n)}(\xi)$ satisfy the recurrence 
relations
\begin{eqnarray}
\a^{(0)}\,_{s}A_{lm}^{(1)}(\xi)+\b^{(0)}\,_{s}A_{lm}^{(0)}(\xi)&=&0, 
\label{eq:3termElm0}\\
\a^{(n)}\,_{s}A_{lm}^{(n+1)}(\xi)+\b^{(n)}\,_{s}A_{lm}^{(n)}(\xi)
+\c^{(n)}\,_{s}A_{lm}^{(n-1)}(\xi)&=&0, \, (n\ge 1),
\label{eq:3termElm}
\end{eqnarray}
where
\begin{eqnarray}
\a^{(n)}&=&
\frac{4\xi(n+\alpha+1)(n+\beta+1)(n+(\alpha+\beta)/2+1-s)}{(2n+\alpha+\beta+2)
(2n+\alpha+\beta+3)},\nonumber \\
\b^{(n)}&=&
\,_{s}E_{l}^{m}(\xi)+\xi ^2-\left(n+\frac{\alpha+\beta}{2}\right)
\left(n+\frac{\alpha+\beta}{2}+1\right)\nonumber \\
&&+\frac{2\xi s(\alpha-\beta)(\alpha+\beta)}{(2n+\alpha+\beta)(2n+\alpha+\beta+2)},\nonumber \\
\c^{(n)}&=&
-\frac{4\xi n(n+\alpha+\beta)(n+(\alpha+\beta)/2+s)}{(2n+\alpha+\beta-1)(2n+
\alpha+\beta)}.
\end{eqnarray}

The method to determine the eigenvalue $\,{_s}E_{l}^{m}$ is similar to
that in the case of renormalized angular momentum presented in $\S$ \ref{sec:about nu}. 
The three-term recurrence relation Eq. (\ref{eq:3termElm}) 
has two independent solutions, which behave for large $n$ as
\begin{eqnarray}
&&A_{(1)}^{(n)}\sim \frac{{\rm const.}(-\xi)^n}
{\Gamma(n+(\alpha+\beta+3)/2-s)}, \label{eq:AlmMin}\\
&&A_{(2)}^{(n)}\sim {\rm const.}\xi^n \Gamma(n+(\alpha+\beta+1)/2+s).
\label{eq:AlmDom}
\end{eqnarray}
The first one, $A_{(1)}^{(n)}$, is the minimal solution, and the 
second one, $A_{(2)}^{(n)}$, is a dominant solution, since
$\lim_{n\rightarrow \infty}A_{(1)}^{(n)}/A_{(2)}^{(n)}=0$. 
For the case of $A_{(2)}^{(n)}$, these coefficients increase with $n$, 
and the series Eq. (\ref{eq:Jacobi-series}) diverges for all values of $x$. 
In the case of $A_{(1)}^{(n)}$, this series 
converges. Thus, we have to choose $A_{(1)}^{(n)}$ in the series expansion
Eq. (\ref{eq:Jacobi-series}). 

Next, we introduce the following:
\begin{equation}
R_n\equiv {A_{(1)}^{n}\over A_{(1)}^{n-1}},\quad
L_n\equiv {A_{(1)}^{n}\over A_{(1)}^{n+1}}.
\end{equation}
The ratio $R_n$ can be expressed as a continued fraction:
\begin{equation}
R_n=\frac{A_{(1)}^{(n)}}{A_{(1)}^{(n-1)}}=
-{\gamma^{(n)}\over \beta^{(n)}-}
{\alpha^{(n)}\gamma^{(n+1)}\over \beta^{(n+1)}-}
{\alpha^{(n+1)}\gamma^{(n+2)}\over \beta^{(n+2)}-}\cdots . 
\label{eq:RncontElm}
\end{equation}
We can also express $L_n$ as a continued fraction:
\begin{eqnarray}
L_n&=&-{\alpha^{(n)}\over {\beta^{(n)}+\gamma^{(n)} L_{n-1}}}
\nonumber\\
&=&-{\alpha^{(n)}\over \beta^{(n)}-}\,
{\alpha^{(n-1)}\gamma^{(n)}\over \beta^{(n-1)}-}\,
{\alpha^{(n-2)}\gamma^{(n-1)}\over \beta^{(n-2)}-}\cdots
{\alpha^{(1)}\gamma^{(2)}\over \beta^{(1)}-}\,
{\alpha^{(0)}\gamma^{(1)}\over \beta^{(0)}}.
\label{eq:LncontElm}
\end{eqnarray}
This expression for $R_n$ is valid if the continued fraction 
converges. By using the properties of the three-term recurrence relations 
(Ref. \citen{Gautschi}, p. 35), it is proved that the continued fractions 
Eq. (\ref{eq:RncontElm}) converge as long as the 
eigenvalue $\,_{s}E_{l}^{m}$ is finite. 

Now, we divide Eq. (\ref{eq:3termElm}) by the expansion coefficients 
$\,_{s}A_{lm}^{(n)}$ and obtain 
\begin{eqnarray}
h(\,_{s}E_{\ell}^{m})\equiv \b^{(n)}+\a^{(n)}R_{n+1}+\c^{(n)}L_{n-1}=0.
\label{eq:determine_elm}
\end{eqnarray}
We replace $R_{n+1}$ and $L_{n-1}$ by the continued fractions given in 
Eqs. (\ref{eq:RncontElm}) and (\ref{eq:LncontElm}). 
We can determine the eigenvalue $\,_{s}E_{\ell}^{m}$ as a root of 
the equation Eq. (\ref{eq:determine_elm}), 
where $l=n+(\a+\b)/2$. Because the equations in Eq. (\ref{eq:determine_elm}) with 
different values of $n$ are independent, 
they give different values of $\,_{s}E_{\ell}^{m}$. 
We thus obtain each $\,_{s}E_{\ell}^{m}$ for different $\ell$ even if 
the recurrence relation Eq. (\ref{eq:3termElm}) does not contain $\ell$
explicitly.

As in $\S$ \ref{sec:about nu}, we adopt {\rm Brent's algorithm}\cite{Recipes} 
in order to determine $_{s}E_{\ell}^m$. In the case that
$a\omega$ is not large, we can use an analytical expression of 
$_{s}E_{\ell}^m$ as the initial value of the root search algorithm.
An analytical expression of $_{s}E_{\ell}^m$ in a series of powers of 
$\xi=a\omega$ is given by
\begin{eqnarray}
_{s}E_{\ell}^m = \ell (\ell +1) -\frac{2 s^2 m}{\ell (\ell +1)} \xi 
	+ \left[H(\ell+1)-H(\ell)-1\right]\xi^2 +O(\xi^3),
\end{eqnarray}
where 
\begin{eqnarray}
H(\ell)=\frac{2(\ell^2-m^2)(\ell^2-s^2)^2}{(2\ell-1)\ell^3(2\ell+1)}.
\end{eqnarray}
The function $h(\,_{s}E_{\ell}^{m})$ is usually monotonic, 
and therefore it is very easy to find the root of Eq. (\ref{eq:determine_elm}).

After we obtain the eigenvalue $_{s}E_{\ell}^m$, 
we can determine all the coefficients. 
The coefficient for $n=n_\ell=l-(\alpha+\beta)/2$ is usually the 
largest term. The ratio of other terms to those for $n=n_\ell$, i.e.
$A_{(1)}^{(n)}/A_{(1)}^{(n_{\ell})}$, 
can be determined using Eqs. (\ref{eq:RncontElm}) and (\ref{eq:LncontElm})
from $n=n_\ell$ to $n=0$ or $n=\infty$. 

The factor $A_{(1)}^{(n_{\ell})}$ is determined 
by the normalization condition. 
For this, we introduce a new 
function $\,_{s}V_{lm}(x)$ through
\begin{eqnarray}
\label{eq:SpheV}
_{s}S_{lm}^{a\omega}(x)=e^{-\xi x}\left(\frac{1-x}{2}\right)^{\frac{\alpha}{2}}
\left(\frac{1+x}{2}\right)^{\frac{\beta}{2}}\, _{s}V_{lm}(x).
\end{eqnarray}
From Eqs. (\ref{eq:SpheU}) and (\ref{eq:SpheV}), we find
\begin{eqnarray}
\label{eq:sphUtoV}
\,_{s}V_{lm}(x)={\rm exp}(2\xi x)\,_{s}U_{lm}(x).
\end{eqnarray}
We insert Eq. (\ref{eq:SpheV}) into Eq. (\ref{eq:Sphe diff}) and find 
that $\,_{s}V_{lm}(x)$ satisfies the differential equation
\begin{eqnarray}
\label{eq:proto-JacobiV}
&&(1-x^{2})\,_{s}V_{lm}''(x)+\left[\beta-\alpha-(2+\alpha+\beta)x\right]\,
_{s}V_{lm}'(x)
\nonumber\\
&&\quad
+\left[\,_{s}E_{l}^{m}(\xi)-\frac{\alpha+\beta}{2}\left(\frac{\alpha+\beta}{2}+
1\right)\right]\,_{s}V_{lm}(x)\nonumber \\
&=&\xi\left[2(1-x^{2})\,_{s}V_{lm}'(x)-(\alpha+\beta-2s+2)x\,_{s}V_{lm}(x)
\right.
\nonumber\\
&&\quad
\left.
-(\xi-\beta+\alpha)\,_{s}V_{lm}(x)\right].
\end{eqnarray}

In the same way as in the case of $_{s}U_{lm}(x)$, 
we expand $_{s}V_{lm}(x)$ in a series of Jacobi polynomials: 
\begin{eqnarray}
\label{eq:Jacobi-series2}
_{s}V_{lm}(x)=\sum_{n=0}^{\infty}\,_{s}B_{lm}^{(n)} \,P_{n}^{(\alpha,\beta)}(x).
\end{eqnarray}
The expansion coefficients $_{s}B_{lm}^{(n)}(\xi)$ satisfy the recurrence 
relations
\begin{eqnarray}
\tilde{\a}^{(0)}\,_{s}B_{lm}^{(1)}(\xi)+\tilde{\b}^{(0)}\,_{s}B_{lm}^{(0)}(\xi)
&=&0, \nonumber \\
\tilde{\a}^{(n)}\,_{s}B_{lm}^{(n+1)}(\xi)+\tilde{\b}^{(n)}\,_{s}B_{lm}^{(n)}
(\xi)+\tilde{\c}^{(n)}\,_{s}B_{lm}^{(n-1)}(\xi)&=&0, \quad (n\ge 1)
\label{eq:3termElm2}
\end{eqnarray}
where
\begin{eqnarray}
\tilde{\a}^{(n)}&=&
-\frac{4\xi(n+\alpha+1)(n+\beta+1)(n+(\alpha+\beta)/2+1+s)}{(2n+\alpha+\beta+2)
(2n+\alpha+\beta+3)},\nonumber \\
\tilde{\b}^{(n)}&=&
\,_{s}E_{l}^{m}(\xi)+\xi ^2-\left(n+\frac{\alpha+\beta}{2}\right)
\left(n+\frac{\alpha+\beta}{2}+1\right)\nonumber \\
&&+\frac{2\xi s(\alpha-\beta)(\alpha+\beta)}{(2n+\alpha+\beta)(2n+\alpha+\beta+2)},\nonumber \\
\tilde{\c}^{(n)}&=&
\frac{4\xi n(n+\alpha+\beta)(n+(\alpha+\beta)/2-s)}{(2n+\alpha+\beta-1)(2n+
\alpha+\beta)}.
\end{eqnarray}

In order for
the series Eq. (\ref{eq:Jacobi-series2}) to converge, the coefficients $_{s}B_{lm}^{(n)}$
must constitute the minimal solution of the recurrence relation Eq. (\ref{eq:3termElm2}).
Suppose $\{B_{(1)}^{(n)}\}$ is the minimal solution. Then, we have
\begin{eqnarray}
\frac{B_{(1)}^{(n)}}{B_{(1)}^{(n-1)}}&=&
-{\tilde{\gamma}^{(n)}\over \tilde{\beta}^{(n)}-}\,
{\tilde{\alpha}^{(n)}\tilde{\gamma}^{(n+1)}\over \tilde{\beta}^{(n+1)}-}\,
{\tilde{\alpha}^{(n+1)}\tilde{\gamma}^{(n+2)}\over \tilde{\beta}^{(n+2)}-}\cdots , 
\label{eq:RncontElm2}
\\
\frac{B_{(1)}^{(n)}}{B_{(1)}^{(n+1)}}&=&
-{\tilde{\alpha}^{(n)}\over \tilde{\beta}^{(n)}-}\,
{\tilde{\alpha}^{(n-1)}\tilde{\gamma}^{(n)}\over \tilde{\beta}^{(n-1)}-}\,
{\tilde{\alpha}^{(n-2)}\tilde{\gamma}^{(n-1)}\over \tilde{\beta}^{(n-2)}-}\cdots
{\tilde{\alpha}^{(1)}\tilde{\gamma}^{(2)}\over \tilde{\beta}^{(1)}-}\,
{\tilde{\alpha}^{(0)}\tilde{\gamma}^{(1)}\over \tilde{\beta}^{(0)}}.
\label{eq:LncontElm2}
\end{eqnarray}
From these equations, we can determine the ratios of 
all the coefficients, $B_{(1)}^{(n)}/B_{(1)}^{(n_{\ell})}$. 

Now, we determine the value of the two coefficients
$A_{(1)}^{(n_{\ell})}$ and $B_{(1)}^{(n_{\ell})}$.
Because Eq. (\ref{eq:sphUtoV}) must be satisfied for any value of $x$, 
we set $x=1$ in Eq. (\ref{eq:sphUtoV}) and obtain
\begin{eqnarray}
\,_{s}B_{lm}^{(n_{\ell})}(\xi)\sum_{n=0}^{\infty}\frac{\,_{s}B_{lm}^{(n)}(\xi)}
{\,_{s}B_{lm}^{(n_{\ell})}(\xi)}\binom{n+\a}{n}=
{\rm exp}(2\xi)\,_{s}A_{lm}^{(n_{\ell})}(\xi)\sum_{n=0}^{\infty}
\frac{\,_{s}A_{lm}^{(n)}(\xi)}{\,_{s}A_{lm}^{(n_{\ell})}(\xi)}\binom{n+\a}{n}.
\nonumber\\
\label{eq:normalization1}
\end{eqnarray}
Also, from the normalization condition Eq. (\ref{eq:normalSp}), we find
\begin{eqnarray}
\int_{-1}^{1}{\rm d}x\left(\frac{1-x}{2}\right)^{\a}\left(\frac{1+x}{2}\right)
^{\b}\sum_{n_{1}=0}^{\infty}\,_{s}A_{lm}^{(n_{1})}P_{n_{1}}^{(\a,\b)}(x)
\sum_{n_{2}=0}^{\infty}\,_{s}B_{lm}^{(n_{2})}P_{n_{2}}^{(\a,\b)}(x)=1.
\nonumber\\
\label{eq:AlmBlm}
\end{eqnarray}

\begin{table}[htbp]
\begin{center}
\caption{$E_{lm}$ for $Mw=0.1$.}
\begin{tabular}{cc|cc}
\hline \hline 
$\ell$&$m$&$q=-0.9$&$q=0.9$\\ \hline 
$2$&$2$&$6.2340859091\times 10^{0}$&$5.7540002160\times 10^{0}$\\
$2$&$1$&$6.1153623140\times 10^{0}$&$5.8752937655\times 10^{0}$\\
$2$&$0$&$5.9957562220\times 10^{0}$&$5.9957562220\times 10^{0}$\\
$2$&$-1$&$5.8752937655\times 10^{0}$&$6.1153623140\times 10^{0}$\\
$2$&$-2$&$5.7540002160\times 10^{0}$&$6.2340859091\times 10^{0}$\\
$3$&$3$&$1.2175986227\times 10^{1}$&$1.1815913335\times 10^{1}$\\
$3$&$2$&$1.2116698236\times 10^{1}$&$1.1876700620\times 10^{1}$\\
$3$&$1$&$1.2057141841\times 10^{1}$&$1.1937158320\times 10^{1}$\\
$3$&$0$&$1.1997300442\times 10^{1}$&$1.1997300442\times 10^{1}$\\
$3$&$-1$&$1.1937158320\times 10^{1}$&$1.2057141841\times 10^{1}$\\
$3$&$-2$&$1.1876700620\times 10^{1}$&$1.2116698236\times 10^{1}$\\
$3$&$-3$&$1.1815913335\times 10^{1}$&$1.2175986227\times 10^{1}$\\
$4$&$4$&$2.0141120497\times 10^{1}$&$1.9853070617\times 10^{1}$\\
$4$&$3$&$2.0105083654\times 10^{1}$&$1.9889090042\times 10^{1}$\\
$4$&$2$&$2.0069068104\times 10^{1}$&$1.9925093222\times 10^{1}$\\
$4$&$1$&$2.0033067556\times 10^{1}$&$1.9961086374\times 10^{1}$\\
$4$&$0$&$1.9997075735\times 10^{1}$&$1.9997075735\times 10^{1}$\\
$4$&$-1$&$1.9961086374\times 10^{1}$&$2.0033067556\times 10^{1}$\\
$4$&$-2$&$1.9925093222\times 10^{1}$&$2.0069068104\times 10^{1}$\\
$4$&$-3$&$1.9889090042\times 10^{1}$&$2.0105083654\times 10^{1}$\\
$4$&$-4$&$1.9853070617\times 10^{1}$&$2.0141120497\times 10^{1}$\\
$5$&$5$&$3.0117824408\times 10^{1}$&$2.9877790649\times 10^{1}$\\
$5$&$4$&$3.0093445069\times 10^{1}$&$2.9901445845\times 10^{1}$\\
$5$&$3$&$3.0069155402\times 10^{1}$&$2.9925172191\times 10^{1}$\\
$5$&$2$&$3.0044953129\times 10^{1}$&$2.9948972039\times 10^{1}$\\
$5$&$1$&$3.0020835961\times 10^{1}$&$2.9972847731\times 10^{1}$\\
$5$&$0$&$2.9996801598\times 10^{1}$&$2.9996801598\times 10^{1}$\\
$5$&$-1$&$2.9972847731\times 10^{1}$&$3.0020835961\times 10^{1}$\\
$5$&$-2$&$2.9948972039\times 10^{1}$&$3.0044953129\times 10^{1}$\\
$5$&$-3$&$2.9925172191\times 10^{1}$&$3.0069155402\times 10^{1}$\\
$5$&$-4$&$2.9901445845\times 10^{1}$&$3.0093445069\times 10^{1}$\\
$5$&$-5$&$2.9877790649\times 10^{1}$&$3.0117824408\times 10^{1}$\\
$6$&$6$&$4.2101144835\times 10^{1}$&$4.1895407126\times 10^{1}$\\
$6$&$5$&$4.2083478965\times 10^{1}$&$4.1912048117\times 10^{1}$\\
$6$&$4$&$4.2065910934\times 10^{1}$&$4.1928777542\times 10^{1}$\\
$6$&$3$&$4.2048439822\times 10^{1}$&$4.1945596361\times 10^{1}$\\
$6$&$2$&$4.2031064705\times 10^{1}$&$4.1962505533\times 10^{1}$\\
$6$&$1$&$4.2013784654\times 10^{1}$&$4.1979506008\times 10^{1}$\\
$6$&$0$&$4.1996598734\times 10^{1}$&$4.1996598734\times 10^{1}$\\
$6$&$-1$&$4.1979506008\times 10^{1}$&$4.2013784654\times 10^{1}$\\
$6$&$-2$&$4.1962505533\times 10^{1}$&$4.2031064705\times 10^{1}$\\
$6$&$-3$&$4.1945596361\times 10^{1}$&$4.2048439822\times 10^{1}$\\
$6$&$-4$&$4.1928777542\times 10^{1}$&$4.2065910934\times 10^{1}$\\
$6$&$-5$&$4.1912048117\times 10^{1}$&$4.2083478965\times 10^{1}$\\
$6$&$-6$&$4.1895407126\times 10^{1}$&$4.2101144835\times 10^{1}$\\

\hline \hline
\end{tabular}
\label{tab:elm}
\end{center}
\end{table}

\begin{table}[htbp]
\begin{center}
\caption{Spin weighted spheroidal harmonics  for $M\omega=0.1$ and $\cos\theta=-0.9$.}
\begin{tabular}{cc|rr}
\hline \hline 
$\ell$&$m$&$q=-0.9$&$q=0.9$\\ \hline 
$2$&$2$&$4.3386981105\times 10^{-3}$&$3.5950865567\times 10^{-3}$\\
$2$&$1$&$-3.7087357243\times 10^{-2}$&$-3.1984590567\times 10^{-2}$\\
$2$&$0$&$1.9411700165\times 10^{-1}$&$1.7424257031\times 10^{-1}$\\
$2$&$-1$&$-6.7728189805\times 10^{-1}$&$-6.3276040875\times 10^{-1}$\\
$2$&$-2$&$1.4469520014\times 10^{0}$&$1.4070158595\times 10^{0}$\\
$3$&$3$&$2.6579791433\times 10^{-3}$&$2.3432796925\times 10^{-3}$\\
$3$&$2$&$-2.3106765235\times 10^{-2}$&$-2.0900105378\times 10^{-2}$\\
$3$&$1$&$1.2376650604\times 10^{-1}$&$1.1489471197\times 10^{-1}$\\
$3$&$0$&$-4.4858142218\times 10^{-1}$&$-4.2770231715\times 10^{-1}$\\
$3$&$-1$&$1.0509804998\times 10^{0}$&$1.0312667647\times 10^{0}$\\
$3$&$-2$&$-1.1695230417\times 10^{0}$&$-1.1936637613\times 10^{0}$\\
$3$&$-3$&$-9.1785502130\times 10^{-1}$&$-8.8540840083\times 10^{-1}$\\
$4$&$4$&$1.3964986285\times 10^{-3}$&$1.2717039185\times 10^{-3}$\\
$4$&$3$&$-1.2581385396\times 10^{-2}$&$-1.1651738472\times 10^{-2}$\\
$4$&$2$&$7.0862042637\times 10^{-2}$&$6.6759828566\times 10^{-2}$\\
$4$&$1$&$-2.7776127987\times 10^{-1}$&$-2.6634215480\times 10^{-1}$\\
$4$&$0$&$7.5267562259\times 10^{-1}$&$7.3543750538\times 10^{-1}$\\
$4$&$-1$&$-1.2550656615\times 10^{0}$&$-1.2540767509\times 10^{0}$\\
$4$&$-2$&$6.8973736289\times 10^{-1}$&$7.2640111281\times 10^{-1}$\\
$4$&$-3$&$1.2548549058\times 10^{0}$&$1.2430573062\times 10^{0}$\\
$4$&$-4$&$4.9003837955\times 10^{-1}$&$4.7271311939\times 10^{-1}$\\
$5$&$5$&$6.8987908263\times 10^{-4}$&$6.4066680667\times 10^{-4}$\\
$5$&$4$&$-6.4686885018\times 10^{-3}$&$-6.0782911246\times 10^{-3}$\\
$5$&$3$&$3.8436104725\times 10^{-2}$&$3.6551010190\times 10^{-2}$\\
$5$&$2$&$-1.6262489225\times 10^{-1}$&$-1.5656422649\times 10^{-1}$\\
$5$&$1$&$4.9724643493\times 10^{-1}$&$4.8496714304\times 10^{-1}$\\
$5$&$0$&$-1.0442277264\times 10^{0}$&$-1.0333677913\times 10^{0}$\\
$5$&$-1$&$1.2430710500\times 10^{0}$&$1.2553442781\times 10^{0}$\\
$5$&$-2$&$-9.9162402163\times 10^{-2}$&$-1.3327302849\times 10^{-1}$\\
$5$&$-3$&$-1.3250352372\times 10^{0}$&$-1.3300224425\times 10^{0}$\\
$5$&$-4$&$-8.7853110913\times 10^{-1}$&$-8.6299962142\times 10^{-1}$\\
$5$&$-5$&$-2.4419490099\times 10^{-1}$&$-2.3603241859\times 10^{-1}$\\
$6$&$6$&$3.2984531756\times 10^{-4}$&$3.1033107890\times 10^{-4}$\\
$6$&$5$&$-3.2192050631\times 10^{-3}$&$-3.0549982369\times 10^{-3}$\\
$6$&$4$&$2.0138768387\times 10^{-2}$&$1.9279986867\times 10^{-2}$\\
$6$&$3$&$-9.1318426706\times 10^{-2}$&$-8.8215649196\times 10^{-2}$\\
$6$&$2$&$3.0845665000\times 10^{-1}$&$3.0079608194\times 10^{-1}$\\
$6$&$1$&$-7.6021402310\times 10^{-1}$&$-7.4896406387\times 10^{-1}$\\
$6$&$0$&$1.2535400677\times 10^{0}$&$1.2503499550\times 10^{0}$\\
$6$&$-1$&$-1.0079278456\times 10^{0}$&$-1.0282273588\times 10^{0}$\\
$6$&$-2$&$-4.8304664552\times 10^{-1}$&$-4.5779972946\times 10^{-1}$\\
$6$&$-3$&$1.1370699694\times 10^{0}$&$1.1530844521\times 10^{0}$\\
$6$&$-4$&$1.1913495835\times 10^{0}$&$1.1822538661\times 10^{0}$\\
$6$&$-5$&$5.2560056636\times 10^{-1}$&$5.1488724800\times 10^{-1}$\\
$6$&$-6$&$1.1736688454\times 10^{-1}$&$1.1371714705\times 10^{-1}$\\
\hline \hline
\end{tabular}
\label{tab:swsh1}
\end{center}
\end{table}

\begin{table}[htbp]
\begin{center}
\caption{Spin weighted spheroidal harmonics  for $M\omega=0.1$ and $\cos\theta=0$.}
\begin{tabular}{cc|rr}
\hline \hline 
$\ell$&$m$&$q=-0.9$&$q=0.9$\\ \hline 
$2$&$2$&$4.1111211672\times 10^{-1}$&$3.7950484443\times 10^{-1}$\\
$2$&$1$&$-8.0605525927\times 10^{-1}$&$-7.7444328043\times 10^{-1}$\\
$2$&$0$&$9.6770343965\times 10^{-1}$&$9.6770343965\times 10^{-1}$\\
$2$&$-1$&$-7.7444328043\times 10^{-1}$&$-8.0605525927\times 10^{-1}$\\
$2$&$-2$&$3.7950484443\times 10^{-1}$&$4.1111211672\times 10^{-1}$\\
$3$&$3$&$5.8559507510\times 10^{-1}$&$5.5982341302\times 10^{-1}$\\
$3$&$2$&$-9.3732214203\times 10^{-1}$&$-9.3264155844\times 10^{-1}$\\
$3$&$1$&$7.1502176780\times 10^{-1}$&$7.6307833481\times 10^{-1}$\\
$3$&$0$&$3.8419145390\times 10^{-2}$&$-3.8419145390\times 10^{-2}$\\
$3$&$-1$&$-7.6307833481\times 10^{-1}$&$-7.1502176780\times 10^{-1}$\\
$3$&$-2$&$9.3264155844\times 10^{-1}$&$9.3732214203\times 10^{-1}$\\
$3$&$-3$&$-5.5982341302\times 10^{-1}$&$-5.8559507510\times 10^{-1}$\\
$4$&$4$&$7.1154532347\times 10^{-1}$&$6.9134142570\times 10^{-1}$\\
$4$&$3$&$-9.8923696324\times 10^{-1}$&$-9.9458839545\times 10^{-1}$\\
$4$&$2$&$5.0542053337\times 10^{-1}$&$5.5505006446\times 10^{-1}$\\
$4$&$1$&$4.0134847826\times 10^{-1}$&$3.4802911340\times 10^{-1}$\\
$4$&$0$&$-8.3798585128\times 10^{-1}$&$-8.3798585128\times 10^{-1}$\\
$4$&$-1$&$3.4802911340\times 10^{-1}$&$4.0134847826\times 10^{-1}$\\
$4$&$-2$&$5.5505006446\times 10^{-1}$&$5.0542053337\times 10^{-1}$\\
$4$&$-3$&$-9.9458839545\times 10^{-1}$&$-9.8923696324\times 10^{-1}$\\
$4$&$-4$&$6.9134142570\times 10^{-1}$&$7.1154532347\times 10^{-1}$\\
$5$&$5$&$8.1075684145\times 10^{-1}$&$7.9470010304\times 10^{-1}$\\
$5$&$4$&$-1.0107000628\times 10^{0}$&$-1.0200385934\times 10^{0}$\\
$5$&$3$&$3.3710130936\times 10^{-1}$&$3.8104293488\times 10^{-1}$\\
$5$&$2$&$6.0312055648\times 10^{-1}$&$5.6888007421\times 10^{-1}$\\
$5$&$1$&$-7.6566919046\times 10^{-1}$&$-7.8489902785\times 10^{-1}$\\
$5$&$0$&$-2.5485570450\times 10^{-2}$&$2.5485570450\times 10^{-2}$\\
$5$&$-1$&$7.8489902785\times 10^{-1}$&$7.6566919046\times 10^{-1}$\\
$5$&$-2$&$-5.6888007421\times 10^{-1}$&$-6.0312055648\times 10^{-1}$\\
$5$&$-3$&$-3.8104293488\times 10^{-1}$&$-3.3710130936\times 10^{-1}$\\
$5$&$-4$&$1.0200385934\times 10^{0}$&$1.0107000628\times 10^{0}$\\
$5$&$-5$&$-7.9470010304\times 10^{-1}$&$-8.1075684145\times 10^{-1}$\\
$6$&$6$&$8.9272963150\times 10^{-1}$&$8.7970600210\times 10^{-1}$\\
$6$&$5$&$-1.0179015837\times 10^{0}$&$-1.0287597521\times 10^{0}$\\
$6$&$4$&$1.9938800690\times 10^{-1}$&$2.3714685226\times 10^{-1}$\\
$6$&$3$&$7.2764306367\times 10^{-1}$&$7.0598308423\times 10^{-1}$\\
$6$&$2$&$-6.6414145769\times 10^{-1}$&$-6.8995712931\times 10^{-1}$\\
$6$&$1$&$-2.7229278396\times 10^{-1}$&$-2.3136795981\times 10^{-1}$\\
$6$&$0$&$8.1612099685\times 10^{-1}$&$8.1612099685\times 10^{-1}$\\
$6$&$-1$&$-2.3136795981\times 10^{-1}$&$-2.7229278396\times 10^{-1}$\\
$6$&$-2$&$-6.8995712931\times 10^{-1}$&$-6.6414145769\times 10^{-1}$\\
$6$&$-3$&$7.0598308423\times 10^{-1}$&$7.2764306367\times 10^{-1}$\\
$6$&$-4$&$2.3714685226\times 10^{-1}$&$1.9938800690\times 10^{-1}$\\
$6$&$-5$&$-1.0287597521\times 10^{0}$&$-1.0179015837\times 10^{0}$\\
$6$&$-6$&$8.7970600210\times 10^{-1}$&$8.9272963150\times 10^{-1}$\\

\hline \hline
\end{tabular}
\label{tab:swsh2}
\end{center}
\end{table}

\begin{table}[htbp]
\begin{center}
\caption{Spin weighted spheroidal harmonics  for $M\omega=0.1$ and $\cos\theta=0.9$.}
\begin{tabular}{cc|rr}
\hline \hline 
$\ell$&$m$&$q=-0.9$&$q=0.9$\\ \hline 
$2$&$2$&$1.4070158595\times 10^{0}$&$1.4469520014\times 10^{0}$\\
$2$&$1$&$-6.3276040875\times 10^{-1}$&$-6.7728189805\times 10^{-1}$\\
$2$&$0$&$1.7424257031\times 10^{-1}$&$1.9411700165\times 10^{-1}$\\
$2$&$-1$&$-3.1984590567\times 10^{-2}$&$-3.7087357243\times 10^{-2}$\\
$2$&$-2$&$3.5950865567\times 10^{-3}$&$4.3386981105\times 10^{-3}$\\
$3$&$3$&$8.8540840083\times 10^{-1}$&$9.1785502130\times 10^{-1}$\\
$3$&$2$&$1.1936637613\times 10^{0}$&$1.1695230417\times 10^{0}$\\
$3$&$1$&$-1.0312667647\times 10^{0}$&$-1.0509804998\times 10^{0}$\\
$3$&$0$&$4.2770231715\times 10^{-1}$&$4.4858142218\times 10^{-1}$\\
$3$&$-1$&$-1.1489471197\times 10^{-1}$&$-1.2376650604\times 10^{-1}$\\
$3$&$-2$&$2.0900105378\times 10^{-2}$&$2.3106765235\times 10^{-2}$\\
$3$&$-3$&$-2.3432796925\times 10^{-3}$&$-2.6579791433\times 10^{-3}$\\
$4$&$4$&$4.7271311939\times 10^{-1}$&$4.9003837955\times 10^{-1}$\\
$4$&$3$&$1.2430573062\times 10^{0}$&$1.2548549058\times 10^{0}$\\
$4$&$2$&$7.2640111281\times 10^{-1}$&$6.8973736289\times 10^{-1}$\\
$4$&$1$&$-1.2540767509\times 10^{0}$&$-1.2550656615\times 10^{0}$\\
$4$&$0$&$7.3543750538\times 10^{-1}$&$7.5267562259\times 10^{-1}$\\
$4$&$-1$&$-2.6634215480\times 10^{-1}$&$-2.7776127987\times 10^{-1}$\\
$4$&$-2$&$6.6759828566\times 10^{-2}$&$7.0862042637\times 10^{-2}$\\
$4$&$-3$&$-1.1651738472\times 10^{-2}$&$-1.2581385396\times 10^{-2}$\\
$4$&$-4$&$1.2717039185\times 10^{-3}$&$1.3964986285\times 10^{-3}$\\
$5$&$5$&$2.3603241859\times 10^{-1}$&$2.4419490099\times 10^{-1}$\\
$5$&$4$&$8.6299962142\times 10^{-1}$&$8.7853110913\times 10^{-1}$\\
$5$&$3$&$1.3300224425\times 10^{0}$&$1.3250352372\times 10^{0}$\\
$5$&$2$&$1.3327302849\times 10^{-1}$&$9.9162402163\times 10^{-2}$\\
$5$&$1$&$-1.2553442781\times 10^{0}$&$-1.2430710500\times 10^{0}$\\
$5$&$0$&$1.0333677913\times 10^{0}$&$1.0442277264\times 10^{0}$\\
$5$&$-1$&$-4.8496714304\times 10^{-1}$&$-4.9724643493\times 10^{-1}$\\
$5$&$-2$&$1.5656422649\times 10^{-1}$&$1.6262489225\times 10^{-1}$\\
$5$&$-3$&$-3.6551010190\times 10^{-2}$&$-3.8436104725\times 10^{-2}$\\
$5$&$-4$&$6.0782911246\times 10^{-3}$&$6.4686885018\times 10^{-3}$\\
$5$&$-5$&$-6.4066680667\times 10^{-4}$&$-6.8987908263\times 10^{-4}$\\
$6$&$6$&$1.1371714705\times 10^{-1}$&$1.1736688454\times 10^{-1}$\\
$6$&$5$&$5.1488724800\times 10^{-1}$&$5.2560056636\times 10^{-1}$\\
$6$&$4$&$1.1822538661\times 10^{0}$&$1.1913495835\times 10^{0}$\\
$6$&$3$&$1.1530844521\times 10^{0}$&$1.1370699694\times 10^{0}$\\
$6$&$2$&$-4.5779972946\times 10^{-1}$&$-4.8304664552\times 10^{-1}$\\
$6$&$1$&$-1.0282273588\times 10^{0}$&$-1.0079278456\times 10^{0}$\\
$6$&$0$&$1.2503499550\times 10^{0}$&$1.2535400677\times 10^{0}$\\
$6$&$-1$&$-7.4896406387\times 10^{-1}$&$-7.6021402310\times 10^{-1}$\\
$6$&$-2$&$3.0079608194\times 10^{-1}$&$3.0845665000\times 10^{-1}$\\
$6$&$-3$&$-8.8215649196\times 10^{-2}$&$-9.1318426706\times 10^{-2}$\\
$6$&$-4$&$1.9279986867\times 10^{-2}$&$2.0138768387\times 10^{-2}$\\
$6$&$-5$&$-3.0549982369\times 10^{-3}$&$-3.2192050631\times 10^{-3}$\\
$6$&$-6$&$3.1033107890\times 10^{-4}$&$3.2984531756\times 10^{-4}$\\
\hline \hline
\end{tabular}
\label{tab:swsh3}
\end{center}
\end{table}

Because the Jacobi polynomials are orthogonal, we have 
\begin{eqnarray}
&&\int_{-1}^{1}{\rm d}x\left(\frac{1-x}{2}\right)^{\a}\left(\frac{1+x}{2}\right)
^{\b}P_{n_{1}}^{(\a,\b)}(x)P_{n_{2}}^{(\a,\b)}(x)
\nonumber\\
&&\quad
=\frac{2\, \Gamma(n+\a+1)\Gamma(n+\b+1)\delta_{n_{1},n_{2}}}{(2n+\a+\b+1)\Gamma(n+1)\Gamma(n+\a+\b+1)}.
\end{eqnarray}
Then, Eq. (\ref{eq:AlmBlm}) reduces to 
\begin{eqnarray}
&&\sum_{n=0}^{\infty}\left[\frac{\,_{s}A_{lm}^{(n)}}{\,_{s}A_{lm}^{(n_{\ell})}}\right]
\left[\frac{\,_{s}B_{lm}^{(n)}}{\,_{s}B_{lm}^{(n_{\ell})}}\right]
\frac{2\, \Gamma(n+\a+1)\Gamma(n+\b+1)}{(2n+\a+\b+1)\Gamma(n+1)
\Gamma(n+\a+\b+1)}
\nonumber\\
&&\quad
=\frac{1}{\,_{s}A_{lm}^{(n_{\ell})}\,_{s}B_{lm}^{(n_{\ell})}}.
\label{eq:normalization2}
\end{eqnarray}
We can obtain the squares of $\,_{s}A_{lm}^{(n_{\ell})}$ and $\,_{s}B_{lm}^{(n_{\ell})}$ 
from Eqs. (\ref{eq:normalization1}) and (\ref{eq:normalization2}). 
Finally, we determine the signs of $\,_{s}A_{lm}^{(n_{\ell})}$ and $\,_{s}B_{lm}^{(n_{\ell})}$
by requiring that the sign of $_{s}S_{\ell m}^{a\omega}(x)$ 
be the same as the sign of the spin-weight spherical harmonics 
in the limit $a\omega\rightarrow 0$. 

In Table \ref{tab:elm},
we list the eigenvalues of the spin-weighted 
spheroidal harmonics for $q=\pm 0.9$ from $\ell=2$ to $6$
in the case $M\omega=0.1$. 
In Tables \ref{tab:swsh1}--\ref{tab:swsh3}, we list the values of the spin-weighted spheroidal 
harmonics for $\cos\theta=-0.9,0,0.9$, $q=\pm 0.9$ and $\ell=2-6$.


\begin{thebibliography}{99}
\bibitem{GK2} This data was calculated by D.~Kennefick based on his previous work\cite{GK}, 
and was kindly provided for us.
\end{thebibliography}
\end{document}